\documentclass[letterpaper,10pt]{article}

\usepackage{jcappub}
\usepackage{amsmath}
\usepackage{amssymb}
\usepackage{amsfonts}
\usepackage{latexsym}
\usepackage{graphicx}% Include figure files
\usepackage{dcolumn}% Align table columns on decimal point
\usepackage{bm}% bold math
\usepackage{empheq}
\usepackage{hyperref}
\usepackage{color}
\usepackage{times}  %%  Only change the main text font to Times New Roman

\newcommand{\ba}{\begin{eqnarray}}
\newcommand\be{\begin{equation}}
\newcommand\beq{\begin{equation}}
\newcommand\ee{\end{equation}}
\newcommand\eeq{\end{equation}}

\definecolor{gxhighlight}{rgb}{1,1,0.4}
\title{Inflation and primordial non-Gaussianities of ``generalized Galileons''}

\author[a,b,c]{Xian Gao}

\author[a]{Dani\`{e}le A. Steer}%

\affiliation[a]{\href{http://www.apc.univ-paris7.fr/APC_CS/en}{Astroparticule et Cosmologie (APC)},\\
     UMR 7164-CNRS, Universit\'{e} Denis Diderot-Paris 7,
        10 rue Alice Domon et L\'{e}onie Duquet, 75205 Paris, France}
\affiliation[b]{\href{http://www.lpt.ens.fr/?lang=en}{Laboratoire de Physique Th\'{e}orique, \'{E}cole Normale Sup\'{e}rieure (LPTENS)},\\
    24 rue Lhomond, 75231 Paris, France}
\affiliation[c]{\href{http://www.iap.fr/english/}{Institut d'Astrophysique de Paris (IAP)},\\
    UMR 7095-CNRS, Universit\'{e} Pierre et Marie Curie-Paris 6, 98bis Boulevard Arago, 75014 Paris, France.}

\emailAdd{xgao@apc.univ-paris7.fr}
\emailAdd{steer@apc.univ-paris7.fr}

\arxivnumber{1107.2642}
\keywords{modified gravity, inflation, physics of the early universe}%Use showkeys class option if keyword
\abstract{
We set up cosmological perturbation theory and study the cosmological implications
of the so-called ``generalized Galileon'' developed in \cite{Deffayet:2011gz,horndeski}.
This is the most general scalar field theory whose Lagrangian contains derivatives up to second order while keeping second order equations of motion, and contains as sub-cases $k$-inflation, $G$-inflation and many other models. We calculate the power spectrum of the primordial curvature perturbation, finding a modification of the usual consistency relation of the tensor-to-scalar ratio in $k$-inflation or perfect fluid models. Finally we also calculate the bispectrum, which contains no new shapes beyond those of $k$-inflation.
}

\begin{document}
\maketitle%
%\tableofcontents%

%%
%%%%%%%%%%%%%%%%%%%%%%%%%%%%%%%%%%%%%%%%%%%%%%%%%%%%%%%%%%%%%%%%%%%%
%%--------------------    Main Article.Begin    ------------------%%
%%%%%%%%%%%%%%%%%%%%%%%%%%%%%%%%%%%%%%%%%%%%%%%%%%%%%%%%%%%%%%%%%%%%

\section{Introduction}

Recently a great deal of effort has been devoted to developing consistent modified gravity theories which, as an alternative to to dark energy or the cosmological constant, may provide an explanation of cosmic acceleration (for a recent review, see \cite{Clifton:2011jh}). Such IR modification of gravity arise, for example, in higher-dimensional setups as in the DGP model where self-acceleration is sourced by a scalar field $\phi$, the helicity-0 mode of the 5D graviton. On small scales the DGP model also reproduces general relativity due to non-linear interactions operating through the Vainstein mechanism \cite{Vainshtein:1972sx}.
Other theories of modified gravity, such as DBI-galileons can also be obtained from a higher dimensional approach \cite{brane_ga}.

Generally, modified gravity models contain additional degrees of freedom which can often be viewed as scalar fields.
In this paper we consider models with a {\it single} scalar field $\phi$, though allowing for couplings between $\phi$ and the metric which may be very different from the standard couplings and potentials considered in many inflationary models.   Indeed, the context of our work is that of the ``generalized Galileon'' derived in \cite{Deffayet:2011gz} (see also \cite{horndeski}).

While there are numerous different models of inflation with different actions and potentials \cite{inf_model} -- for example chaotic inflation, small/large-field inflation, $k$-inflation and DBI-inflation to name a few -- single-field $k$-inflation \cite{kess}, in particular, attracted a lot of attention since it is the most general scalar field theory with a Lagrangian containing derivatives up to \emph{first} order $\mathcal{L}=\mathcal{L}(\phi,\nabla\phi)$ (and hence, clearly, having second order equations of motion).  Indeed, all the inflationary models listed above are a special case of $k$-inflation, and for that reason the development of  cosmological perturbation theory in general case of $k$-inflation has a broad range of applicability.

In this paper we go beyond $k$-inflation and study the most general Lagrangian in 4 space-time dimensions which contain both first {\it and second} derivatives of the scalar field, $\mathcal{L}=\mathcal{L}(\phi,\nabla\phi,\nabla\nabla\phi)$, but constructed  such a way that the field equations for both the metric and the scalar field remain second order \cite{horndeski,Deffayet:2011gz}. This prevents the theory from having extra degrees of freedom
as well as Ostrogradski instabilities \cite{Ostrogradski},  and thus yields a possibly viable higher-order derivative scalar field theory. This general Lagrangian includes, in certain limits, the decoupling limit of DGP model \cite{Dvali:2000hr} as well as some consistent
theories of massive gravity \cite{dgpmass,deRham:2011by}; and it also includes the ``Galileon'' model
\cite{Nicolis:2008in}, the conformal Galileon \cite{conf_gal}, K-mouflage \cite{Babichev:2009ee}, and also ``$G$-'' or ``KGB'' inflation \cite{Deffayet:2010qz,g-inf}. Finally, in a very simple limit (which can be useful to check results) it also reduces to $k$-inflation.
Our aim is to study cosmological perturbation theory in the context of this very general Lagrangian, outlined in Section \ref{sec:model}, working both to second as well as third order so as to study non-Gaussianities. The formalism we develop is therefore applicable to a large class of models. As such, this paper is rather technical -- since we crank the standard handle of cosmological perturbation theory -- but the results are rather general.

As we will see below, see equation (\ref{action}), the Lagrangian we consider \cite{Deffayet:2011gz} contains {\it four free functions} which will play an important r\^ole in the following. We denote them by
\be
K(X,\phi),\qquad \text{and}\qquad G^{(n)}(X,\phi), \qquad n=1,2,3
\ee
where $\phi$ is the scalar field, and
\be
X=-\frac{1}{2}g^{\mu \nu}\partial_\mu \, \phi \partial_\nu \phi,
\label{Xdef}
\ee
is its kinetic term
(we work with a mostly positive signature).  To obtain the Lagrangian of $k$-inflation \cite{kess} one must simply set
\be
K\neq 0 \; ,  \qquad G^{(1,2,3)}(X,\phi) = 0, \qquad \qquad (\text{$k$-inflation limit}).
\label{kinflation-limit}
\ee
The  Galileon Lagrangian \cite{Nicolis:2008in,cdcov} is obtained when
\be
K = X c_{(0)},\qquad   G^{(1)} = Xc_{(1)},\qquad G_{,X}^{(2,3)}=Xc_{(2,3)},  \qquad \qquad ({\rm Galileon \; limit})
\label{Galileon-limit}
\ee
where the $c_{(n)}$ are (dimensionful) constants.
(The effective scalar field Lagrangian in the decoupling limit of the DGP model is obtained by setting $c_{(2)}=c_{(3)} = 0$.)
In this case the Lagrangian will be invariant under shifts of the field, namely $\phi\rightarrow \phi+c$.
In general, when $K(X,\phi)$ and $G^{(n)}(X,\phi)$ are non-vanishing functions of $X$ and $\phi$, shift symmetry is broken (as indeed is the ``Galileon'' symmetry \cite{Nicolis:2008in} since we work in curved backgrounds). Without shift symmetry, questions of the stability of the theory against large renormalisation may arise \cite{bdst_ng}, but we do not dwell on this point here.

To obtain the Lagrangian of $G$-inflation \cite{g-inf} (or equivalently KGB-inflation \cite{Deffayet:2010qz}), one must set
\be
K,  G^{(1)} \neq 0 , \qquad G^{(2)}=  G^{(3)} = 0 \; \qquad \qquad \text{($G$-inflation limit}).
\label{Ginflation-limit}
\ee
Cosmological perturbations has been studied in Galileon model \cite{defelice,kobayashi,g-inf,lss_gal,galileon_inf,kyy2,Charmousis:2011bf,Gubitosi:2011sg}. Primordial non-Gaussianities \cite{defe_ng,mk_ng,bdst_ng,cdmnt_ng,kyy1,renaux} have been studied in $G$-inflation limit, and it was shown
\cite{cdmnt_ng} that to leading order in slow-roll parameters there are no new momentum shapes for the bispectrum (beyond the standard ones of $k$-inflation). Despite that, a distinctive feature of $G$-inflation is that, though it has only a single scalar field, its energy momentum tensor takes the form of an imperfect  fluid \cite{Deffayet:2010qz,Pujolas:2011he} (contrary to the $k$-essence models).  This fact sheds some light on the understanding of imperfect fluids in cosmology, and furthermore can lead to violations of the null enery condition thus opening up a whole range of possibly exotic phenomena \cite{cnt}.  Problems of superluminality may arise in this case, though it is still under debate as to whether causality is violated in general \cite{Babichev:2007dw}. Previous understanding of the conservation of curvature perturbation on super-Hubble scales relied highly on $k$-essence or perfect fluid models: it is interesting to see \cite{Naruko:2011zk,Gao:2011mz} that this conservation law still holds in the generalized Galileon model, that is when the functions $K$ and $G^{(n)}$ are all taken to be arbitrary functions of $\phi$ and $X$.

In this work, we keep the greatest generality by taking the full Lagrangian of generalized Galileon \cite{Deffayet:2011gz,horndeski} into account: that is, arbitrary functions $K$ and $G^{(n)}$. We derive the background equations of motion for an inflationary background, around which we calculate the power spectrum and the bispectrum of the primordial curvature perturbation $\zeta$ (the background evolution and linear perturbations of the generalized Galileon \cite{Deffayet:2011gz,horndeski} were also investigated very recently in \cite{kyy2}, see also \cite{Charmousis:2011bf} for a discussion of self-tuning mechanism on FRW background). We find a modification of the familiar consistency relation of the tensor-to-scalar ratio.
From the theoretical point of view, the results presented here can be used to study the evolution of cosmological perturbations as well as the observational implications numerous models.

This paper is organized as follows. In Sec.~\ref{sec:model}, we review the generalized Galileon model. Background evolution of our model and linear perturbations are studied in Sec.~\ref{sec:linear}. In Sec.~\ref{sec:bispectrum}, we compute the full third-order action for the curvature perturbation and evaluate the corresponding contributions to the bispectrum. Final section is devoted to conclusion and discussion.

%%%%%%%%%%%%%%%%%%%%%%%
\section{Generalized Galileons: Field equations}{\label{sec:model}}

In \cite{Deffayet:2011gz}, the most general scalar field action which contains at most second derivatives of $\phi$, is polynomial in these second derivatives, and which leads to second order equations of motion in flat space-time, was constructed in $D$ dimensional space-time. Its ``covariatization" \cite{cdcov} leads to an action with field equations which are second order or lower for both $\phi$ and the metric. In 4D this is the unique action with second order equations of motion \cite{horndeski}, and is given by
    \begin{equation}{\label{action}}
        S = \int d^4x \sqrt{-g} \left( \frac{1}{2}R + \sum_{n=0}^{3}\mathcal{L}_{n} \right),
    \end{equation}
where we have set $M_{\text{pl}}^2=1/(8\pi G)=1$, and
    \begin{eqnarray}
        \mathcal{L}^{(0)} & = & K^{}\left(X,\phi\right),\\
        \mathcal{L}^{(1)} & = & G^{(1)}\left(X,\phi\right)\Box\phi,\\
        \mathcal{L}^{(2)} & = & G_{,X}^{(2)}\left(X,\phi\right)\left[\left(\Box\phi\right)^{2}-\left(\nabla_{\mu}\nabla_{\nu}\phi\right)^{2}\right]
        +R\, G^{(2)}\left(X,\phi\right) ,\\
        \mathcal{L}^{(3)} & = & G_{,X}^{(3)}\left(X,\phi\right)\left[\left(\Box\phi\right)^{3}-3\Box\phi\left(\nabla_{\mu}\nabla_{\nu}\phi\right)^{2}
        +2\left(\nabla_{\mu}\nabla_{\nu}\phi\right)^{3}\right]-6G_{\mu\nu}\nabla^{\mu}\nabla^{\nu}\phi\,
        G^{(3)}\left(X,\phi\right) .
        \end{eqnarray}
Here, as discussed in the introduction, $K(X,\phi)$ and $G^{(n)}(X,\phi)$ are arbitrary functions of $\phi$ and $X$ (defined in (\ref{Xdef})).
In $\mathcal{L}^{(2)}$, $\left(\nabla_{\mu}\nabla_{\nu}\phi\right)^{2} = \left(\nabla_{\mu}\nabla_{\nu}\phi\right)\left(\nabla^{\mu}\nabla^{\nu}\phi\right)$ and $R$ is the Ricci scalar. Thus in principle the Einstein-Hilbert action $R/2$ in (\ref{action}) can be viewed as a special case of $\mathcal{L}^{(2)}$ with $G^{(2)}=1/2$. However, since for certain applications of our analysis we may wish to set $\mathcal{L}^{(2)}=0$ (see (\ref{kinflation-limit})-(\ref{Ginflation-limit})), we keep the Einstein-Hilbert term explicit.  Whilst the Lagrangian $\mathcal{L}^{(0)}$ and $\mathcal{L}^{(1)}$ are familiar from $k$ and $G$-inflation, terms $\mathcal{L}^{(2,3)}$ represent completely new class of scalar field theories which has not been explored so far.  In $\mathcal{L}^{(3)}$,
$$
\left(\nabla_{\mu}\nabla_{\nu}\phi\right)^{3} =  \left(\nabla_{\mu}\nabla_{\nu}\phi\right)\left(\nabla^{\mu}\nabla^{\rho}\phi\right)\left(\nabla_{\rho}\nabla^{\nu}\phi\right)
$$
whereas $G_{\mu \nu}$ is the Einstein tensor. Notice that a term $G_{\mu \nu} \partial^\mu \phi \partial^\nu \phi$, discussed in for example \cite{Gubitosi:2011sg}, can by obtained by setting $G^{(3)} \propto \phi$.

For completeness, we collect the corresponding equations of motion for the Galileon scalar field and the metric. The Einstein equation following from (\ref{action}) is
    \begin{equation}
        G_{\mu\nu} = T_{\mu\nu} \equiv \sum_{n=0}^{3} T^{(n)}_{\mu\nu},
        \label{Einst}
    \end{equation}
where a superscript ``${}^{(n)}$" denotes the contribution from $\mathcal{L}^{(n)}$. Using the notation $G^{(n)}_{,X} = \partial G^{(n)}/\partial X$ and $G^{(n)}_{,\phi} = \partial G^{(n)}/\partial \phi$, the stress-energy tensors $T_{\mu\nu}^{(0,1)}$ are given by \cite{Deffayet:2010qz,Pujolas:2011he}
    \begin{eqnarray}
        T_{\mu\nu}^{(0)} & = & K^{}g_{\mu\nu}+K_{,X}^{}\nabla_{\mu}\phi\nabla_{\nu}\phi,\label{EMT0}\\
        T_{\mu\nu}^{(1)} & = & - \left(\nabla_{\lambda} G^{(1)}  \nabla^{\lambda}\phi \right) g_{\mu\nu} + 2 \nabla_{(\mu} G^{(1)} \nabla_{\nu)} \phi + \square\phi G_{,X}^{(1)} \nabla_{\mu}\phi\nabla_{\nu}\phi
        \nonumber
        \\
        &=& -\left(G_{,X}^{(1)}\nabla_{\lambda}X \nabla^{\lambda}\phi-2XG_{,\phi}^{(1)}\right)g_{\mu\nu}
%         \nonumber\\
%        &&
        +\left(G_{,X}^{(1)}\square\phi+2G_{,\phi}^{(1)}\right)\nabla_{\mu}\phi\nabla_{\nu}\phi+2G_{,X}^{(1)}\nabla_{(\mu}\phi\nabla_{\nu)}X. \label{EMT1}
        \end{eqnarray}
Here, from (\ref{Xdef}), $\nabla_{\mu}X = -\nabla_{\mu}\nabla_{\lambda}\phi\nabla^{\lambda}\phi$. As observed in \cite{Deffayet:2010qz,Pujolas:2011he}, $T_{\mu\nu}^{(1)}$ contains  second derivatives (but not higher, by construction) of $\phi$, and as a result Einstein's equations (\ref{Einst}) (as well as the scalar field equations of motion) contain second derivatives of {\it both} the metric $g_{\mu \nu}$ {\it and} of $\phi$ as soon as ${\cal L}^{(1)}$ is present.  There is no conformal transformation that can diagonalise the system of equations, which remain coupled: this phenomenon has been dubbed {\it kinetic braiding} \cite{Deffayet:2010qz}. The expressions for $T_{\mu\nu}^{(2,3)}$ are rather more involved, again containing second derivatives of $\phi$, but they now {\it also} contain second derivatives of $g_{\mu \nu}$. In particular \cite{kyy2,Gao:2011mz}
    \begin{eqnarray}
        T_{\mu\nu}^{(2)} & = & g_{\mu\nu}\Big\{ RG^{(2)}-G_{,X}^{(2)}\left(\left(\square\phi\right)^{2}-\left(\nabla_{\rho}\nabla_{\sigma}\phi\right)^{2}\right)+4XG_{,\phi\phi}^{(2)}-2G_{,XX}^{(2)}\nabla_{\rho}X\nabla^{\rho}X\nonumber \\
         &  & -2\left(2G_{,X\phi}^{(2)}+\square\phi G_{,XX}^{(2)}\right)\nabla_{\rho}\phi\nabla^{\rho}X+2\square\phi\left(2XG_{,X\phi}^{(2)}-G_{,\phi}^{(2)}\right)+2G_{,X}^{(2)}R_{\rho\sigma}\nabla^{\rho}\phi\nabla^{\sigma}\phi\Big\}\nonumber \\
         &  & +\left[G_{,X}^{(2)}R+4\square\phi G_{,X\phi}^{(2)}+2G_{,\phi\phi}^{(2)}+G_{,XX}^{(2)}\left(\left(\square\phi\right)^{2}-\left(\nabla_{\rho}\nabla_{\sigma}\phi\right)^{2}\right)\right]\nabla_{\mu}\phi\nabla_{\nu}\phi\nonumber \\
         &  & +4\left(\square\phi G_{,XX}^{(2)}+2G_{,X\phi}^{(2)}\right)\nabla_{(\mu}\phi\nabla_{\nu)}X+2G_{,XX}^{(2)}\left(\nabla_{\mu}X\nabla_{\nu}X-2\nabla_{\rho}X\nabla^{\rho}\nabla_{(\mu}\phi\nabla_{\nu)}\phi\right)\nonumber \\
         &  & +2\left(G_{,X}^{(2)}\square\phi+G_{,XX}^{(2)}\nabla_{\rho}\phi\nabla^{\rho}X-2XG_{,X\phi}^{(2)}+G_{,\phi}^{(2)}\right)\nabla_{\mu}\nabla_{\nu}\phi\nonumber \\
         &  & -2G_{,X}^{(2)}\left(\nabla^{\rho}\nabla_{\mu}\phi\nabla_{\nu}\nabla_{\rho}\phi+2\nabla_{(\mu}\phi R_{\nu)\rho}\nabla^{\rho}\phi+R_{\rho\mu\sigma\nu}\nabla^{\rho}\phi\nabla^{\sigma}\phi\right)-2G^{(2)}R_{\mu\nu}.\label{EMT2}
        \end{eqnarray}
The expression for $T^{(3)}_{\mu \nu}$ is given in Appendix \ref{sec:EMT3}.

While the stress-energy tensor $T^{(0)}_{\mu\nu}$ is clearly of the perfect fluid form, $T_{\mu\nu}^{(1,2,3)}$ are not. In the local rest frame defined by the effective four-velocity
\be
u_\mu = \frac{\nabla_\mu \phi}{\sqrt{2X}} \qquad u_\mu u^\mu =-1
\ee
the stress energy tensors take the general form
\be
T_{\mu\nu}^{(n)} = \rho^{(n)} u_\mu u_\nu + P^{(n)} (g_{\mu \nu } + u_\mu u_\nu) +   \left( u_{\mu}q^{(n)}_{\nu}  + u_{\nu}q^{(n)}_{\mu} \right) + \pi^{(n)}_{\mu \nu}
\ee
where $\rho^{(n)} = T_{\mu\nu}^{(n)} u^\mu u^\nu$ is the energy density, $P^{(n)}$ the isotropic pressure, $q^{(n)}_{\mu}$ the energy flow, and $\pi^{(n)}_{\nu \mu}=\pi^{(n)}_{\mu \nu}$ the anisotropic stress.  The expressions for $\rho^{(1)}$, $P^{(1)}$ and $q^{(1)}$ can straightforwardly be read off from (\ref{EMT1}) from which it follows that $\pi^{(1)}_{\mu \nu}=0$.  For $T_{\mu\nu}^{(2,3)}$ the situation is yet more complicated, and from (\ref{EMT2}) and (\ref{EMT3}) they both lead to non-vanishing anisotropic stress $\pi^{(2,3)}_{\mu \nu}$ as well as energy flow.
The consequences of this non-perfect fluid picture have been discussed in \cite{Deffayet:2010qz,Pujolas:2011he,Gao:2011mz}.

The equation of motion for the scalar field $\phi$ is given by
 \begin{equation}{\label{scalar_eom}}
        \sum_{n=0}^{3}\mathcal{E}^{(n)} = 0,
    \end{equation}
where
\be
\mathcal{E}^{(n)}=\nabla_{\mu}J^{(n)\mu}+\mathcal{L}_{,\phi}^{(n)}.
\label{eofmJ}
\ee
The currents $J^{(n)}$, which are only conserved when the theory is shift-symmetric $ \mathcal{L}_{,\phi}^{(n)}=0$, are given by
\begin{eqnarray}
J^{(0)\mu} & = & K_{,X}^{}\nabla^{\mu}\phi,\\
J^{(1)\mu} & = & \square\phi G_{,X}^{(1)}\nabla^{\mu}\phi+\nabla^{\mu}G^{(1)},\\
J^{(2)\mu} & = & \left[\left(\left(\square\phi\right)^{2}-\left(\nabla_{\mu}\nabla_{\nu}\phi\right)^{2}\right)G_{,XX}^{(2)}+RG_{,X}^{(2)}\right]\nabla^{\mu}\phi+2\nabla_{\nu}\left(G_{,X}^{(2)}\left(\Box\phi g^{\mu\nu}-\nabla^{\mu}\nabla^{\nu}\phi\right)\right),\\
J^{(3)\mu} & = & -\left[6G_{\rho\sigma}\nabla^{\rho}\nabla^{\sigma}\phi G_{,X}^{(3)}-\left(\left(\square\phi\right)^{3}-3\square\phi\left(\nabla_{\rho}\nabla_{\sigma}\phi\right)^{2}+2\left(\nabla_{\rho}\nabla_{\sigma}\phi\right)^{3}\right)G_{,XX}^{(3)}\right]\nabla^{\mu}\phi
\nonumber
\\
 &  & +\nabla_{\nu}\left[G_{,X}^{(3)}\left(3\left(\Box\phi\right)^{2}g^{\mu\nu}-3g^{\mu\nu}\left(\nabla_{\rho}\nabla_{\sigma}\phi\right)^{2}-6\Box\phi\nabla^{\mu}\nabla^{\nu}\phi+6\nabla^{\mu}\nabla_{\lambda}\phi\nabla^{\lambda}\nabla^{\nu}\phi\right)\right] \nonumber
 \\
 &  & -6G^{\mu\nu}\nabla_{\nu}G^{(3)}.
\end{eqnarray}
Though at first sight it might appear that these equations of motion contain derivatives of order 3 or higher, this is not the case: by construction, these higher order derivatives cancel on calculating $\nabla_{\mu}J^{(n)\mu}$. The explicit equations of motion given in Appendix  \ref{sec:eom3}.

\section{Background equations}

We now consider a spatially flat FLRW geometry with metric
\be
ds^2 =a^2(\eta) \left( -d\eta^2 + d\bm{x}^2 \right),
\ee
where $\eta$ is conformal time. Then the kinetic term $X$ reduces to
\be
X = \frac{1}{2a^2} \phi'^2
\ee
where a dash denotes a derivative with respect to $\eta$.   Then the background equations of motion (which most straightfowardly are obtained by the requiring first order perturbative action to be vanishing to vanish) are given by
    \begin{equation}\label{eom_alpha}
        \mathcal{H}^{2}= \frac{a^{2}}{3} \rho
    \end{equation}
and
    \begin{equation}{\label{eom_zeta}}
      \mathcal{H}^{2}+2\mathcal{H}'+a^{2} P = 0,
    \end{equation}
where the effective energy density $\rho$ and pressure $P$ can be read off from (\ref{EMT0}-\ref{EMT2}) as well as (\ref{EMT3}) and are given by
    \begin{equation}
        \rho = \sum_{n=0}^{3}\rho^{(n)},\qquad\qquad P=\sum_{n=0}^{3}P^{(n)},
    \end{equation}
with
    \begin{eqnarray}
\rho^{(0)} & = & 2XK_{,X}^{}-K^{},\label{rho0}\\
\rho^{(1)} & = & -6\frac{\mathcal{H}\phi'}{a^{2}}XG_{,X}^{(1)}+2XG_{,\phi}^{(1)},\label{rho1}\\
\rho^{(2)} & = & -6\frac{\mathcal{H}^{2}}{a^{2}}\left(G^{(2)}-4X\left(G_{,X}^{(2)}+XG_{,XX}^{(2)}\right)\right)-6\frac{\mathcal{H}\phi'}{a^{2}}\left(2XG_{,X\phi}^{(2)}+G_{,\phi}^{(2)}\right),\label{rho2}\\
\rho^{(3)} & = & 36\frac{\mathcal{H}^{2}}{a^{2}}X\left(2XG_{,X\phi}^{(3)}+3G_{,\phi}^{(3)}\right)-\frac{12\mathcal{H}^{3}\phi'}{a^{4}}X\left(5G_{,X}^{(3)}+2XG_{,XX}^{(3)}\right),\label{rho3}
\end{eqnarray}
and
    \begin{eqnarray}
        P^{(0)} & = & K^{},\label{P0}\\
        P^{(1)} & = & 2XG_{,\phi}^{(1)}+\frac{2}{a^{2}}XG_{,X}^{(1)}\left(\phi''-\mathcal{H}\phi'\right),\label{P1}\\
        P^{(2)} & = & 4XG_{,\phi\phi}^{(2)}+\frac{2}{a^{2}}G^{(2)}\left(\mathcal{H}^{2}+2\mathcal{H}'\right)+\frac{4}{a^{2}}XG_{,X\phi}^{(2)}\left(\phi''-3\mathcal{H}\phi'\right)+\frac{2}{a^{2}}G_{,\phi}^{(2)}\left(\mathcal{H}\phi'+\phi''\right)\nonumber \\
         &  & +\frac{8}{a^{2}}\mathcal{H}XG_{,XX}^{(2)}\left(2\mathcal{H}X-\frac{\phi'\phi''}{a^{2}}\right)+\frac{1}{a^{2}}G_{,X}^{(2)}\left[4X\left(\mathcal{H}^{2}-2\mathcal{H}'\right)-\frac{4\mathcal{H}}{a^{2}}\phi'\phi''\right],\label{P2}\\
        P^{(3)} & = & -\frac{24}{a^{2}}\mathcal{H}XG_{,\phi\phi}^{(3)}\phi'+\frac{24\mathcal{H}^{2}}{a^{4}}X^{2}G_{,XX}^{(3)}\left(\phi''-\mathcal{H}\phi'\right)+\frac{12\mathcal{H}}{a^{4}}G_{,X}^{(3)}X\left[\left(2\mathcal{H}'-3\mathcal{H}^{2}\right)\phi'+3\mathcal{H}\phi''\right]\nonumber \\
         &  & +\frac{24}{a^{2}}\mathcal{H}XG_{,X\phi}^{(3)}\left(3\mathcal{H}X-\frac{\phi'\phi''}{a^{2}}\right)+\frac{12}{a^{2}}G_{,\phi}^{(3)}\left[X\left(3\mathcal{H}^{2}-2\mathcal{H}'\right)-\frac{2\mathcal{H}}{a^{2}}\phi'\phi''\right].\label{P3}
        \end{eqnarray}
Of course the energy fluxes and anisotropic stresses vanish on the background, and (\ref{rho0})-(\ref{P3}) reduce to the standard expressions \cite{kess,Nicolis:2008in,cdcov,g-inf} in the $k$-essence (\ref{kinflation-limit}), Galileon (\ref{Galileon-limit}) and $G$-inflation limits (\ref{Ginflation-limit}) respectively.

The equation of motion for the scalar field follows from (\ref{scalar_eom})  (or alternatively as a combination of (\ref{eom_alpha}) and (\ref{eom_zeta}))
and is given by
\[
\frac{1}{a^{4}}\left(a^{2}J\right)'=\mathcal{L}_{,\phi}
\]
with
\begin{eqnarray*}
J & = & K_{,X}^{}\phi'-\frac{\phi'}{a^{2}}\left(-a^{2}G_{,\phi}^{(1)}+3\mathcal{H}G_{,X}^{(1)}\phi'\right)-\frac{6\mathcal{H}\phi'}{a^{2}}\left(-\mathcal{H}G_{,X}^{(2)}+\phi'\left(G_{,X\phi}^{(2)}-G_{,XX}^{(2)}\frac{\mathcal{H}\phi'}{a^{2}}\right)\right)\\
 &  & -\frac{6\mathcal{H}^{2}\phi'}{a^{6}}\left[-3a^{4}G_{,\phi}^{(3)}+\phi'\left(3a^{2}\mathcal{H}G_{,X}^{(3)}+\phi'\left(-3a^{2}G_{,X\phi}^{(3)}+\mathcal{H}G_{,XX}^{(3)}\phi'\right)\right)\right],
\end{eqnarray*}
and
\begin{eqnarray*}
\mathcal{L}_{,\phi} & = & K_{,\phi}^{}-\frac{1}{a^{2}}G_{,\phi}^{(1)}\left(2\mathcal{H}\phi'+\phi''\right)+\frac{1}{a^{4}}6\left(a^{2}G_{,\phi}^{(2)}\left(\mathcal{H}^{2}+\mathcal{H}'\right)+\mathcal{H}G_{,X\phi}^{(2)}\phi'\phi''\right)\\
 &  & -\frac{6\mathcal{H}}{a^{6}}\left[\mathcal{H}G_{,X\phi}^{(3)}\left(\phi'\right)^{2}\left(-2\mathcal{H}\phi'+3\phi''\right)+3a^{2}G_{,\phi}^{(3)}\left(2\mathcal{H}'\phi'+\mathcal{H}\phi''\right)\right].
\end{eqnarray*}

As usual, we define the slow-roll parameter
 \begin{equation}
        \epsilon \equiv -\frac{d \ln (\mathcal{H}/a)}{d \ln a} = 1-\frac{\mathcal{H}'}{\mathcal{H}^2}
    \end{equation}
which, on using the background equations (\ref{eom_alpha})-(\ref{eom_zeta}), and in order to get a exponential expansion phase, imposes     \begin{equation}
        |\epsilon|=\frac{a^{2}}{2\mathcal{H}^{2}}\left|\rho+P\right|\ll 1.
    \end{equation}
(This generalizes the slow-roll condition for the scalar field in $k$-essence model $\left| \frac{\phi'^{2}}{2\mathcal{H}^{2}}K_{,X}^{} \right| \ll 1$.)

%%%%%%%%%%%%%%%%%%%%%%%%%%%%%%%%%%%%%%%
\section{Linear perturbations}{\label{sec:linear}}

We now derive the second order action governing the dynamics of linear perturbations around the background solution for the general action given in (\ref{action}), using standard techniques (and in particular following \cite{Maldacena:2002vr}).  One should note that this second order action has already been studied in \cite{kyy2}: taking into account differences of notation, we have checked that all our results agree with the corresponding expressions given in \cite{kyy2}.

Before proceeding, it is useful to count the number of dynamical degree(s) of freedom in our system. Since the Lagrangian (\ref{action}) is constructed such that the corresponding field equations of motion (for both the scalar field and the metric) are second-order, there are no additional degree of freedom:
the dynamical degrees of freedom of the model are exactly those of standard General Relativity with minimally coupled single scalar field, namely one propagating scalar mode and two tensor modes.

Given that the curvature perturbation $\zeta$ is conserved on large scales \cite{Gao:2011mz}, we have found it more convenient to carry out perturbation calculations in the uniform scalar field gauge with $\delta\phi=0$.  Then the perturbed metric
takes the form{\footnote{Throughout this paper, spatial indices are raised and lowered and summarized by $\delta_{ij}$ and $\partial^2\beta=\delta^{ij}\partial_i\partial_j\beta$, $\delta^{ij}$, and $(\partial_i\beta)^2=\delta^{ij}\partial_i\beta\partial_j\beta$, $(\partial_i\partial_j\beta)^2=\delta^{ik}\delta^{jl}\partial_i\partial_j\beta \partial_k\partial_l\beta$ etc should be understood.}}
\begin{equation}
    ds^{2}=a^{2}(\eta)\left[-\left(e^{2\alpha}-e^{-2\zeta}\delta^{ij}\partial_{i}\beta\partial_{j}\beta\right)d\eta^{2}+2\partial_{i}\beta d\eta dx^{i}+e^{2\zeta}\delta_{ij}dx^{i}dx^{j}\right],
    \label{perturbedm}
\end{equation}
so that in terms of ADM variables, the lapse, shift, and metric on spacial slices are given by
\be
N=ae^{\alpha}\; , N_i =a^2\partial_{i}\beta \; \; \; {\rm and} \; \; \;  h_{ij} =a^{2}e^{2\zeta}\delta_{ij}
\ee
respectively.
Just as in minimally coupled $k$-essence, the lapse and shift and thus $\alpha$ and $\beta$ are constraints in our model (\ref{action}). Thus in general $\alpha$, $\beta$ can be solved for perturbatively in terms of curvature perturbation $\zeta$:
    \begin{equation}
        \alpha = \sum_{m=1}\alpha_{(m)},\qquad \beta=\sum_{m=1}\beta_{(m)},
    \end{equation}
where the subscript ``${}_{(m)}$" denotes the order in $\zeta$.  Furthermore, since $\alpha$, $\beta$ are constraints, their first order solution $\alpha_{(1)}$ and $\beta_{(1)}$ are adequate for our purpose to evaluate the power spectrum and bispectrum.

\subsection{Power spectra}
\subsubsection{Scalar perturbations}

On using the form of the perturbed metric (\ref{perturbedm}), a straightforward Taylor expansion of the full action (\ref{action}) together with some integration-by-parts and use of the background equations, yields the quadratic action for $\zeta$, $\alpha$ and $\beta$:
    \begin{equation}{\label{S2_zeta_alphabeta}}
        S_{(2)}\left[\zeta,\alpha,\beta\right]=\int d\eta d^{3}xa^{2}\left[-3g_{\zeta}\zeta'^{2}+c_{\zeta}\left(\partial\zeta\right)^{2}- 3\mathcal{H}^{2}m_{\alpha}\alpha^{2}+2g_{\zeta}\partial\alpha\cdot\partial\zeta+6\mathcal{H}f_{\alpha}\alpha\zeta'+2g_{\zeta}\zeta'\partial^{2}\beta-2\mathcal{H}f_{\alpha}\alpha\partial^{2}\beta\right].
    \end{equation}
In (\ref{S2_zeta_alphabeta}) four dimensionless coefficients appear (that is, dimensionless when we reinstate the Planck mass $M_{\text{pl}}$), and they are given by:
    \begin{eqnarray}
        g_{\zeta}&=&1-4XG_{,X}^{(2)}+2G^{(2)}-\frac{12}{a^{2}}X\left(a^{2}G_{,\phi}^{(3)}-\mathcal{H}G_{,X}^{(3)}\phi'\right),\label{g_{z}eta}\\c_{\zeta}&=&1+2G^{(2)}+\frac{12}{a^{2}}X\left[G_{,X}^{(3)}\left(\phi''-\mathcal{H}\phi'\right)+a^{2}G_{,\phi}^{(3)}\right],\label{c_{z}eta}\\m_{\alpha}&=&\frac{1}{2}-\frac{a^{2}}{6\mathcal{H}^{2}}\left(K^{}+4X^{2}K_{,XX}^{}\right)-\frac{a^{2}}{3\mathcal{H}^{2}}X\left(2XG_{,X\phi}^{(1)}+G_{,\phi}^{(1)}\right)+\frac{\phi'}{\mathcal{H}}X\left(3G_{,X}^{(1)}+2XG_{,XX}^{(1)}\right)\\&&+G^{(2)}-2X\left(5G_{,X}^{(2)}+2X\left(7G_{,XX}^{(2)}+2XG_{,XXX}^{(2)}\right)\right)+\frac{\phi'}{\mathcal{H}}\left(4X^{2}G_{,XX\phi}^{(2)}+8XG_{,X\phi}^{(2)}+G_{,\phi}^{(2)}\right)\\&&+\frac{2\mathcal{H}\phi'}{a^{2}}X\left(25G_{,X}^{(3)}+4X\left(6G_{,XX}^{(3)}+XG_{,XXX}^{(3)}\right)\right)-6X\left(4X^{2}G_{,XX\phi}^{(3)}+16XG_{,X\phi}^{(3)}+9G_{,\phi}^{(3)}\right),\label{m_{a}lpha}\\f_{\alpha}&=&1+\frac{\phi'}{\mathcal{H}}XG_{,X}^{(1)}+2\left(G^{(2)}-4X\left(G_{,X}^{(2)}+XG_{,XX}^{(2)}\right)\right)+\frac{\phi'}{\mathcal{H}}\left(2XG_{,X\phi}^{(2)}+G_{,\phi}^{(2)}\right)\\&&+\frac{6\mathcal{H}\phi'}{a^{2}}X\left(5G_{,X}^{(3)}+2XG_{,XX}^{(3)}\right)-12X\left(2XG_{,X\phi}^{(3)}+3G_{,\phi}^{(3)}\right).\label{f_{a}lpha}
         \end{eqnarray}
Here $g_{\zeta}$ etc are normalized such that in $k$-essence models ($K^{} \neq 0$, $G^{(1,2,3)} = 0$),
    \begin{eqnarray*}
        g_{\zeta}&=&c_{\zeta}=f_{\alpha}=1,\qquad \text{and} \\ m_{\alpha}&=&1-\frac{a^{2}}{3\mathcal{H}^{2}}X\left(K_{,X}^{}+2XK_{,XX}^{}\right) = 1-\frac{\epsilon}{3c_{s}^{2}},
        \end{eqnarray*}
where
 $c_s^2$ is the standard expression in $k$-essence models, namely $c_s^2 = K_{,X}^{}/\left(K_{,X}^{}+2XK_{,XX}^{}\right)$ which we will derive below. Notice more generally that neither $K^{}$ nor $G^{(1)}$ contribute to $g_\zeta$ and $c_\zeta$: thus in the G-inflation limit we also have $g_\zeta=c_\zeta=1$. On the other hand $G^{(2)}$ and $G^{(3)}$ contribute to all four expressions.  Finally, in the standard Galileon limit (\ref{Galileon-limit}) the above expressions simplify a great deal.

The constraints $\alpha$ and $\beta$ are now obtained by varying (\ref{S2_zeta_alphabeta}), yielding
    \begin{eqnarray}
        \alpha & = & \frac{g_{\zeta}}{\mathcal{H}f_{\alpha}} \zeta',\label{alpha_sol}\\
        \partial^{2}\beta & = & \frac{g_{\zeta}}{\mathcal{H}f_{\alpha}} \partial^{2}\zeta -3\left(\frac{g_{\zeta}m_{\alpha}}{f_{\alpha}^{2}}-1\right)\zeta'.\label{beta_sol}
        \end{eqnarray}
(Again one can verify that in $k$-essence limit, (\ref{alpha_sol})-(\ref{beta_sol}) reduces to the standard results \cite{Chen:2006nt} $\alpha = \frac{\zeta'}{\mathcal{H}}$ and $\partial^2\beta = -\frac{\partial^{2}\zeta}{\mathcal{H}}+\frac{a^{2}}{\mathcal{H}^{2}}X\left(K_{,X}^{}+2XK_{,XX}^{}\right)\zeta' = -\frac{\partial^{2}\zeta}{\mathcal{H}}+\frac{\epsilon}{c_{s}^{2}}\zeta'$.)
On substituting the constraints (\ref{alpha_sol})-(\ref{beta_sol}) solution into (\ref{S2_zeta_alphabeta}), we obtain the final quadratic action for $\zeta$,
    \begin{equation}{\label{S2}}
        S_{(2)}[\zeta] = \int d\eta d^3x a^2\frac{\epsilon_s}{c_s^2}  \left({\zeta'}^2 - c_s^2 (\partial\zeta)^2\right),
    \end{equation}
which is controlled by two key coefficients
    \begin{eqnarray}
        \epsilon_s &\equiv& \frac{1}{a^{2}}\left(\frac{a^{2}g_{\zeta}^{2}}{\mathcal{H}f_{\alpha}}\right)'-c_{\zeta} , {\label{epsilon_s_def}}\\
        c_s^2 & = & \frac{\epsilon_s}{3g_{\zeta}}\left(1-\frac{g_{\zeta}m_{\alpha}}{f_{\alpha}^{2}}\right)^{-1},\label{cs_def}\end{eqnarray}
which determine the normalization and propagation speed of $\zeta$ respectively.  In the $k$-inflation limit, $\epsilon_s = \epsilon$ whereas $c_s^2$ is given by the familiar expression \cite{kess}. In $G$-inflation, our expressions agree with those of \cite{Deffayet:2010qz,g-inf} (on using the background equations of motion), while in the pure Galileon limit they agree with those given \cite{bdst_ng}.  Finally, similar expressions for the general Galileon model can be found in \cite{kobayashi}.

Finally, in order to avoid ghost and instabilities, we must impose
\be
\epsilon_s>0 \, ,\qquad c_s^2>0.
\ee
Notice, however, that since $\epsilon_s$ is not simply related to the slow-roll parameter $\epsilon$, requiring $\epsilon_s>0$ does not necessarily impose $\dot{H}<0$.

\subsubsection{Tensor perturbations}

For completeness, we also consider the tensor perturbation
    \begin{equation}
        ds^2 = a^2(\eta)\left[ -d\eta^2 + \left(\delta_{ij} + h_{ij}\right)dx^idx^j  \right],
    \end{equation}
where the symmetric tensor  $h_{ij}$ is transverse and traceless.
The quadratic action for the tensor modes $h_{ij}$ can be easily derived as
    \begin{equation}{\label{S2_tensor}}
        S_{(2)T} = \int d\eta d^3x~ \frac{a^2}{8} \frac{\epsilon_T}{c_T^2}\left[\left(h_{ij}'\right)^{2}-c_{T}^{2}\left(\partial_{k}h_{ij}\right)^{2}\right],
    \end{equation}
where
    \begin{equation}{\label{epsilon_c_T_def}}
        \epsilon_{T} = c_{\zeta}, \qquad\qquad
        c_{T}^2 = \frac{c_{\zeta}}{g_{\zeta}}.
    \end{equation}
Notice that no new parameters enter into the  tensor perturbations, and that in both $k$-inflation and $G$-inflation where $c_\zeta= g_\zeta = 1$, we have $\epsilon_T=c_T=1$.

In the general model (\ref{action}), both the amplitude and the propagating speed of gravitational waves are modified relative to $k$-essence model. In particular,  the tensor perturbation can be either amplified or suppressed by tuning the parameters $g_{\zeta}$ and $c_{\zeta}$, which can be done only with $G^{(2,3)}$.
Finally, in order to avoid ghost and instabilities, we require
    \begin{equation}
        g_{\zeta} >0,\qquad c_{\zeta}>0.
    \end{equation}

We now introduce the polarization decomposition
    \begin{equation}
        h_{ij}(\eta,\bm{k})=\sum_{s=+,\times} h^{s}(\eta,\bm{k})e_{ij}^{s}(k),
    \end{equation}
where the symmetric tensors $e_{ij}^{+,\times}$ are transverse and traceless $k^i\epsilon_{ij}^s(\bm{k})=\epsilon^s_{ii}=0$, and satisfy the  orthogonal and normalization condition:
$e^{s}_{ij} (\bm{k})e^{s'\ast}_{ij} (\bm{k})=2 \delta_{ss'}$. In terms of $h^{s}$,  action (\ref{S2_tensor}) can be rewritten as
    \begin{equation}{\label{S2_tensor_polar}}
        S_{\left(2\right)T} = \sum_{s=+,\times}\int d\eta d^3x\frac{a^{2}}{4}\frac{\epsilon_T}{c_T^2}\left(h_{s}'^{2}-c_{T}^{2}(\partial h_{s})^2\right).
    \end{equation}

\subsubsection{Quantization}

We now quantize the perturbations following the standard procedure e.g. \cite{Mukhanov:1990me}.

To this end, we assume a quasi-de Sitter inflationary background, and that the Hubble parameter $H$, $\epsilon_s$, $c_s$, $\epsilon_T$ and $c_T$ are slowly varying with time during the period when cosmological perturbations, whose scales are of the current observational interest, are generated and exit the Hubble scales. In this case, we use the familiar mathematical trick, i.e. first evaluating the power spectra by treating $H$, $\epsilon_s$, $c_s$, $\epsilon_T$ and $c_T$ are exact constant, then taking into account their time-dependence by identifying $H=H(\eta_{\ast}(k))$, $\epsilon_s=\epsilon_s(\eta_{\ast}(k))$ etc with $-c_s k \eta_{\ast} = 1$ in order to evaluating the spectrum indices, up to the first order in slow-varying parameters.

The canonically normalized variable corresponding to $\zeta$ is defined by $\tilde{\zeta}\equiv \frac{a}{c_s}\sqrt{2\epsilon_s}\zeta\equiv z\zeta$, and from (\ref{S2}) it satisfies the familiar equation of motion
    \begin{equation}{\label{eom_tilde_zeta}}
        \tilde{\zeta}_{\bm{k}}'' +\left( c_s^2 k^2 - \frac{z''}{z} \right)\tilde{\zeta}_{\bm{k}} = 0.
    \end{equation}
On selecting positive frequency solutions which correspond to the standard Bunch-Davis vacuum deep inside the Hubble scale, and on converting back to $\zeta$, we find
    \begin{equation}
        \zeta_k(\eta) \simeq \frac{iH}{2\sqrt{\epsilon_s c_{s}k^{3}}}\left(1+ic_{s}k\eta\right)e^{-ic_{s}k\eta},
    \end{equation}
(where $H={\cal H}/a$) so that the power spectrum on super-Hubble scales ($|c_s k\eta| \ll 1$) is given by
    \begin{equation}{\label{P_zeta}}
        \mathcal{P}_{\zeta} =  \frac{H^{2}}{8\pi^{2}\epsilon_s c_{s}},
    \end{equation}
where all quantities are evaluated around the Hubble exit at $|c_s k \eta| =1$.  Notice that this differs from the usual single field result in that $\epsilon$ is replaced by $\epsilon_s$ which, contrary to $k$-inflation, may not necessarily be small (a larger $\epsilon_s$ would tend to decrease the power spectrum).
Similarly, the tensor power spectrum is given by
    \begin{equation}{\label{P_T}}
        \mathcal{P}_T = \frac{2H^{2}}{\pi^{2}\epsilon_{T}c_{T}},
    \end{equation}
which can again be amplified, or reduced, depending on the values of $\epsilon_{T}c_{T}$.
At this point, we emphasize that our expressions (\ref{P_zeta}) and (\ref{P_T}) are consistent with the corresponding results got in \cite{kyy2}.

The tensor-to-scalar ratio therefore reads
    \begin{equation}
        r\equiv \frac{\mathcal{P}_{T}}{\mathcal{P}_{\zeta}} = \frac{16\epsilon_s c_{s}}{\epsilon_{T}c_{T}}.
    \end{equation}
The spectral indices for scalar and tensor perturbations  are given by
    \begin{eqnarray}
        n_{\zeta} -1 & \equiv& \frac{d \ln \mathcal{P}_{\zeta}}{d \ln k} = -2\epsilon - \eta_s - s,\\
        n_{T} & \equiv& \frac{d \ln \mathcal{P}_{T}}{d \ln k} = -2\epsilon - \eta_T - s_T,
    \end{eqnarray}
where in the above,
    \begin{equation}
        \eta_s = \frac{d \ln \epsilon_s}{d \ln a},\qquad \eta_T = \frac{d \ln \epsilon_T}{d \ln a},\qquad s= \frac{d \ln c_s}{d \ln a},\qquad s_T = \frac{d \ln c_T}{d \ln a}.
        \label{def-slow-varying1}
    \end{equation}
In terms of these slow-varying parameters, $\epsilon_s$ defined in (\ref{epsilon_s_def}) can be recast as\footnote{Note that we have not managed to eliminate $f_\alpha$ and use it here as a slowly varying parameter.}
    \begin{equation}
        \epsilon_s =  \frac{1}{f_{\alpha}}\frac{\epsilon_{T}^{2}}{c_{T}^{4}}\left(\epsilon+1+a^{2}\left(2\eta_{T}-s_{f}-4s_{T}\right)\right)-\epsilon_{T},
    \end{equation}
where
\be
s_f = d\ln f_{\alpha}/d \ln a.
\label{def-slow-varying2}
\ee
In $k$-inflation and $G$-inflaton,  $\epsilon_T=c_T=1$, $\eta_T=s_T=0$, $n_T= -2\epsilon$, so that the tensor-to-scalar ratio $r=16\epsilon_s c_s$. In $k$-inflation where $\epsilon_s=\epsilon$, then $r=-8c_s n_T$ which is the standard consistency relation. In $G$-inflation $\epsilon_s$ and $\epsilon$ are related in a complicated manner, but one can again derive a modified consistency relation. This was already found in \cite{defelice,defe_ng,kobayashi,kyy1,kyy2}, and in our case is expressed as
\begin{equation}
        r= -8c_{s}\frac{\epsilon_{T}}{f_{\alpha}c_{T}^{5}}\left[n_{T}+2\left(\frac{f_{\alpha}c_{T}^{4}}{\epsilon_{T}}-1\right)+\eta_{T}+s_{T}-2a^{2}\left(2\eta_{T}-s_{f}-4s_{T}\right)\right].
    \end{equation}
Thus, this gives the first distinctive feature of generalized Galileon model different from $k$-essence model.

%%%%%%%%%%%%%%%%%%%%%%%%%%%%%%%%%%%%%%%%%%%%%%%%%
\section{Bispectrum of curvature perturbation $\zeta$}{\label{sec:bispectrum}}

\subsection{Cubic action for the curvature perturbation}

In this section we again follow the standard approach and extend the previous calculation to third order in perturbation theory.

Due to the
presence of various higher-order derivative terms in generalized Galileon model (\ref{action}), the number of cubic order terms in perturbative Lagrangian is very large!  In order to group various terms, we have proceeded by the following steps. First, we integrate by parts to remove the time derivatives of $\alpha$ and $\beta$, and second we simply (or are able even to eliminate) terms using the background equations of motion (see Appendix \ref{sec:cancel}). After rather tedious and cumbersome multiple integrations-by-parts, we eventually arrive at the following third-order action for $\zeta$, $\alpha$ and $\beta$:
    \begin{eqnarray}
         & & S_{(3)}[\zeta,\alpha,\beta] \nonumber \\
         & = & \int d\eta d^{3}x~ a^{2}\left\{g_{\zeta}\left[-9\zeta\zeta'^{2}+2\zeta'\left(\zeta\partial^{2}\beta+\partial_{i}\zeta\partial^{i}\beta\right)-\alpha\left(\partial_{i}\zeta\right)^{2}+\left(\partial_{i}\beta\right)^{2}\partial^{2}\zeta-\frac{1}{2}\zeta\left(4\alpha\partial^{2}\zeta-\left(\partial^{2}\beta\right)^{2}+\left(\partial_{i}\partial_{j}\beta\right)^{2}\right)\right]\right.\nonumber \\
         &  & +c_{\zeta}\zeta\left(\partial_{i}\zeta\right)^{2}-9\mathcal{H}^{2}m_{\alpha}\alpha^{2}\zeta+2\mathcal{H}f_{\alpha}\alpha\left(9\zeta\zeta'-\zeta\partial^{2}\beta-\partial_{i}\zeta\partial^{i}\beta\right)\nonumber \\
         &  & +\frac{\lambda_{1}}{\mathcal{H}}\left[\zeta'^{3}-\zeta'^{2}\partial^{2}\beta+\frac{1}{2}\zeta'\left(4\alpha\partial^{2}\zeta+\left(\partial^{2}\beta\right)^{2}-\left(\partial_{i}\partial_{j}\beta\right)^{2}\right)-\alpha\left(\partial^{2}\zeta\partial^{2}\beta-\partial_{i}\partial_{j}\zeta\partial_{i}\partial_{j}\beta\right)\right]\nonumber \\
         &  & \left.+\lambda_{2}\alpha\left[3\zeta'^{2}-2\zeta'\partial^{2}\beta+\frac{1}{2}\left(\left(\partial^{2}\beta\right)^{2}-\left(\partial_{i}\partial_{j}\beta\right)^{2}\right)\right]-\lambda_{3}\mathcal{H}\alpha^{2}\left(3\zeta'-\partial^{2}\beta\right)-\lambda_{4}\alpha^{2}\partial^{2}\zeta+\frac{\lambda_{5}}{2}\mathcal{H}^{2}\alpha^{3}\right\}.\label{S3_zeta_alpha_beta}
        \end{eqnarray}
This{\footnote{We have compared our result with (e.g.) the corresponding results recently presented in \cite{defe_ng}. It is interesting to find that, (\ref{S3_zeta_alpha_beta}) has essentially the same structure with eq.(46) in \cite{defe_ng}, which also has 28 cubic terms controlled by nine independent coefficients. More interestingly, these coefficients have exactly the same relations among them.}} contains 28 cubic interaction terms and nine dimensionless coefficients: four of these are $g_{\zeta}$, $c_{\zeta}$, $f_{\alpha}$ and $m_{\alpha}$ which already appeared in the second order action (\ref{S2_zeta_alphabeta}); and the remainder we call
$\lambda_{i},~(i=1,\cdots,5)$, whose explicit expressions are:
    \begin{eqnarray}
\lambda_{1} & = & -\frac{12\mathcal{H}\phi'}{a^{2}}G_{,X}^{(3)}X,\label{lambda1}\\
\lambda_{2} & = & 1+2\left[G^{(2)}-4X\left(G_{,X}^{(2)}+XG_{,XX}^{(2)}\right)\right]+\frac{12\mathcal{H}\phi'}{a^{2}}X\left(5G_{,X}^{(3)}+2XG_{,XX}^{(3)}\right)-12X\left(2XG_{,X\phi}^{(3)}+3G_{,\phi}^{(3)}\right),\label{lambda2}\\
\lambda_{3} & = & 1+\frac{\phi'}{\mathcal{H}}X\left(3G_{,X}^{(1)}+2XG_{,XX}^{(1)}\right)\nonumber \\
 &  & +2\left[G^{(2)}-2X\left(5G_{,X}^{(2)}+2X\left(7G_{,XX}^{(2)}+2XG_{,XXX}^{(2)}\right)\right)\right]+\frac{\phi'}{\mathcal{H}}\left(4X^{2}G_{,XX\phi}^{(2)}+8XG_{,X\phi}^{(2)}+G_{,\phi}^{(2)}\right)\nonumber \\
 &  & +\frac{6\mathcal{H}\phi'}{a^{2}}X\left[25G_{,X}^{(3)}+4X\left(6G_{,XX}^{(3)}+XG_{,XXX}^{(3)}\right)\right]-12X\left(4X\left(XG_{,XX\phi}^{(3)}+4G_{,X\phi}^{(3)}\right)+9G_{,\phi}^{(3)}\right),\label{lambda3}\\
\lambda_{4} & = & 1+2\left(G^{(2)}+4X^{2}G_{,XX}^{(2)}\right)-\frac{12\mathcal{H}\phi'}{a^{2}}X\left(3G_{,X}^{(3)}+2XG_{,XX}^{(3)}\right)+12X\left(2XG_{,X\phi}^{(3)}+G_{,\phi}^{(3)}\right),\label{lambda4}\\
\lambda_{5} & = & 1+\frac{a^{2}}{3\mathcal{H}^{2}}\left[K^{}-2X\left(K_{,X}^{}+6XK_{,XX}^{}+4X^{2}K_{,XXX}^{}\right)\right]\nonumber \\
 &  & +\frac{2\phi'}{\mathcal{H}}X\left[9G_{,X}^{(1)}+4X\left(4G_{,XX}^{(1)}+XG_{,XXX}^{(1)}\right)\right]-\frac{2a^{2}}{3\mathcal{H}^{2}}X\left(4X^{2}G_{,XX\phi}^{(1)}+8XG_{,X\phi}^{(1)}+G_{,\phi}^{(1)}\right)\nonumber \\
 &  & +2\left[G^{(2)}-4X\left(7G_{,X}^{(2)}+4X\left(10G_{,XX}^{(2)}+7XG_{,XXX}^{(2)}+X^{2}G_{,XXXX}^{(2)}\right)\right)\right]\nonumber \\
 &  & +\frac{2\phi'}{\mathcal{H}}\left(8X^{3}G_{,XXX\phi}^{(2)}+36X^{2}G_{,XX\phi}^{(2)}+26XG_{,X\phi}^{(2)}+G_{,\phi}^{(2)}\right)\nonumber \\
 &  & +\frac{4\mathcal{H}\phi'}{a^{2}}X\left(125G_{,X}^{(3)}+218XG_{,XX}^{(3)}+84X^{2}G_{,XXX}^{(3)}+8X^{3}G_{,XXXX}^{(3)}\right)\nonumber \\
 &  & -12X\left(8X^{3}G_{,XXX\phi}^{(3)}+60X^{2}G_{,XX\phi}^{(3)}+98XG_{,X\phi}^{(3)}+27G_{,\phi}^{(3)}\right).\label{lambda5}
\end{eqnarray}
where again, $\lambda_i$ are properly normalized.
%We wish to emphasize that (\ref{S3_zeta_alpha_beta}) is exact --- no approximation was made in its derivation.

At this point, it is useful to recall the $k$-essence model values of $\lambda_i$'s, which are given by
    \begin{eqnarray}
        \lambda_1 &= & 0,\qquad \qquad \lambda_2=\lambda_3=\lambda_4= 1,\\
        \lambda_5 &=& 1+\frac{a^{2}}{3\mathcal{H}^{2}}\left[K^{}-2X\left(K_{,X}^{}+6XK_{,XX}^{}+4X^{2}K_{,XXX}^{}\right)\right].{\label{lambda5_kess}}
    \end{eqnarray}
Using the background equations of motion, in $k$-inflation models, $\lambda_5$ can alternatively be written as $\lambda_5 = -\frac{4a^{2}}{\mathcal{H}^{2}}\lambda$, where
    \[
        \lambda=X^{2}K_{,XX}^{}+\frac{2}{3}X^{3}K_{,XXX}^{},
    \]
which is the ``popular" combination which was introduced (e.g.) in \cite{Chen:2006nt} (note the original definition of $\lambda$ has dimension as $\mathcal{H}^2$). Thus $\lambda_5$ is the natural generalization of $\lambda$ in the Galileon model (\ref{action}), whereas $\lambda_1$, $\lambda_2$, $\lambda_3$ and $\lambda_4$ are new parameters in Galileon model, which are trivial in $k$-essence models. It is interesting to note that in G-inflation, $\lambda_1=0$, $\lambda_2=\lambda_4=1$ as in $k$-essence model. Their corrections arise only when higher order Galileon terms, i.e. $G^{(2)}$ etc are included.

Finally, we can eliminate $\alpha$ and $\beta$  in (\ref{S3_zeta_alpha_beta}) using the constraint solutions (\ref{alpha_sol})-(\ref{beta_sol}). After another set of cumbersome integration-by-parts, the number of cubic interaction terms surprisingly from 28 to 10 and we obtain
    \begin{eqnarray}
        S_{(3)}[\zeta] & = & \int d\eta d^{3}xa^{2}\biggl\{\frac{\Lambda_{1}}{\mathcal{H}}\zeta'^{3}+\Lambda_{2}\zeta'^{2}\zeta+\Lambda_{3}\zeta\left(\partial_{i}\zeta\right)^{2}+\frac{\Lambda_{4}}{\mathcal{H}^{2}}\zeta'^{2}\partial^{2}\zeta+ \Lambda_{5}\zeta'\partial_{i}\zeta\partial^{i}\psi+ \Lambda_{6}\partial^{2}\zeta\left(\partial_{i}\psi\right)^{2}\nonumber \\
         &  & +\frac{\Lambda_{7}}{\mathcal{H}^{2}}\left[\partial^{2}\zeta\left(\partial_{i}\zeta\right)^{2}-\zeta\partial_{i}\partial_{j}\left(\partial^{i}\zeta\partial^{j}\zeta\right)\right] + \frac{\Lambda_{8}}{\mathcal{H}}\left[\partial^{2}\zeta\partial_{i}\zeta\partial^{i}\psi-\zeta\partial_{i}\partial_{j}\left(\partial^{i}\zeta\partial^{j}\psi\right)\right]+F\left(\zeta\right)\left.\frac{\delta{\cal L}_{2}}{\delta\zeta}\right|_{1}\biggr\},\label{S3}
        \end{eqnarray}
with
    \begin{equation}
        \psi \equiv \partial^{-2}\zeta'.
    \end{equation}
The 8 dimensionless parameters in (\ref{S3}) are given by
    \begin{eqnarray}
        \Lambda_{1} & = & \frac{\epsilon_{T}}{c_{T}^{2}}\frac{\epsilon_{s}}{c_{s}^{4}}\frac{1}{f_{\alpha}}+\lambda_{1}\left(1-\frac{\epsilon_{s}c_{T}^{2}}{\epsilon_{T}c_{s}^{2}}\right)+\frac{\lambda_{2}}{f_{\alpha}}\left(3\frac{\epsilon_{T}}{c_{T}^{2}}-2\frac{\epsilon_{s}}{c_{s}^{2}}\right)+\frac{\lambda_{3}}{f_{\alpha}^{2}}\frac{\epsilon_{T}}{c_{T}^{2}}\left(\frac{\epsilon_{s}}{c_{s}^{2}}-3\frac{\epsilon_{T}}{c_{T}^{2}}\right)+\frac{1}{2}\frac{\epsilon_{T}^{3}}{c_{T}^{6}}\frac{\lambda_{5}}{f_{\alpha}^{3}},\label{Lambda1}\\
        \Lambda_{2} & = & \frac{\epsilon_{s}}{c_{s}^{2}}\left[3+\frac{1}{c_{s}^{2}f_{\alpha}}\frac{\epsilon_{T}}{c_{T}^{2}}\left(\epsilon+\eta_{T}-s_f -2s_{T}-3-\eta_{s}\right)\right],\label{Lambda2}\\
        \Lambda_{3} & = & \frac{\epsilon_{T}}{c_{T}^{2}f_{\alpha}}\left[\frac{\epsilon_{s}}{c_{s}^{2}}\left(1+\eta_{T}+\eta_{s}+\epsilon-s_f -2s_{T}-2s\right)+c_{T}^{2}f_{\alpha}-\frac{\epsilon_{T}}{c_{T}^{2}}\left(2\eta_{T}+\epsilon-s_f +1-4s_{T}\right)\right],\label{Lambda3}\\
        \Lambda_{4} & = & \frac{\epsilon_{T}}{c_{T}^{2}f_{\alpha}}\left[3\lambda_{1}-\frac{\epsilon_{T}^{2}}{c_{T}^{4}}\frac{\lambda_{3}}{f_{\alpha}^{2}}+\frac{\epsilon_{T}}{c_{T}^{2}}\left(2\frac{\lambda_{2}}{f_{\alpha}}-\frac{\lambda_{4}}{f_{\alpha}}\right)\right],\label{Lambda4}\\
        \Lambda_{5} & = & -\frac{1}{2}\frac{\epsilon_{s}^{2}c_{T}^{2}}{\epsilon_{T}c_{s}^{4}}\left[1+\lambda_{1}\frac{c_{T}^{2}}{\epsilon_{T}}\left(3-s_{1}-4s_{T}-\epsilon+2\eta_{T}\right)+\frac{\lambda_{2}}{f_{\alpha}}\left(3-2s_{T}-s_{2}-\epsilon+\eta_{T}+s_f \right)\right],\label{Lambda5}\\
        \Lambda_{6} & = & \frac{\epsilon_{s}^{2}c_{T}^{2}}{4\epsilon_{T}c_{s}^{4}}\left[3+\lambda_{1}\frac{c_{T}^{2}}{\epsilon_{T}}\left(\epsilon+s_{1}+4s_{T}-2\eta_{T}-3\right)+\frac{\lambda_{2}}{f_{\alpha}}\left(\epsilon+2s_{T}+s_{2}-\eta_{T}-s_f -3\right)\right],\label{Lambda6}\\
        \Lambda_{7} & = & \frac{\epsilon_{T}^{3}}{6c_{T}^{6}f_{\alpha}^{2}}\left[1+3\lambda_{1}\frac{c_{T}^{2}}{\epsilon_{T}}\left(2\eta_{T}+s_{1}+3\epsilon-1-2s_f -4s_{T}-2\frac{\epsilon_{s}c_{T}^{4}f_{\alpha}}{\epsilon_{T}^{2}}\right)\right.\nonumber \\
         &  & \qquad\qquad\left.+\frac{\lambda_{2}}{f_{\alpha}}\left(3\eta_{T}+s_{2}+3\epsilon-1-6s_{T}-3s_f -3\frac{\epsilon_{s}c_{T}^{4}f_{\alpha}}{\epsilon_{T}^{2}}\right)\right],\label{Lambda7}\\
        \Lambda_{8} & = & -\frac{\epsilon_{s}\epsilon_{T}}{c_{s}^{2}c_{T}^{2}f_{\alpha}}\left[1+\lambda_{1}\frac{c_{T}^{2}}{\epsilon_{T}}\left(s_{1}+2\epsilon-2-s_f -f_{\alpha}\frac{\epsilon_{s}c_{T}^{4}}{\epsilon_{T}^{2}}\right)\right.\nonumber \\
         &  & \qquad\qquad\left.+\frac{1}{2}\frac{\lambda_{2}}{f_{\alpha}}\left(\eta_{T}+s_{2}+2\epsilon-2-2s_{T}-2s_f -2f_{\alpha}\frac{\epsilon_{s}c_{T}^{4}}{\epsilon_{T}^{2}}\right)\right],\label{Lambda8}
        \end{eqnarray}
where $\epsilon_s$, $c_s$, $\epsilon_T$ and $c_T$ are defined in (\ref{epsilon_s_def}), (\ref{cs_def}) and (\ref{epsilon_c_T_def}) respectively, and the slowly varying parameters $\eta_s, \eta_T, s$ and $s_T$ in (\ref{def-slow-varying1}) and (\ref{def-slow-varying2}). Finally
    \begin{equation}
        s_i = \frac{d\ln\lambda_i}{d\ln a}.
    \end{equation}
Equations (\ref{Lambda1})-(\ref{Lambda8}) exactly re-produce  the corresponding values of $\Lambda_i$'s in the $k$-essence limit, which are (recall in $k$-essence limit, $\epsilon_s = \epsilon$, $\epsilon_T=c_T=f_{\alpha}=\lambda_2=\lambda_3=\lambda_4=1$, $\lambda_1=0$, $\lambda_5$ is given in (\ref{lambda5_kess}) and is usually written as $\lambda_5 = -\frac{4\lambda}{{H}^{2}}$)
    \begin{eqnarray*}
\Lambda_{1} & = & \frac{\epsilon}{c_{s}^{2}}\left(\frac{1}{c_{s}^{2}}-1\right)-\frac{2\lambda}{H^{2}},\qquad\qquad
\Lambda_{2}=\frac{\epsilon}{c_{s}^{2}}\left[3+\frac{1}{c_{s}^{2}}\left(\epsilon-3-\eta_{\epsilon}\right)\right],\\
\Lambda_{3} & = & \frac{\epsilon}{c_{s}^{2}}\left(1+\epsilon+\eta_{\epsilon}-2s-c_{s}^{2}\right),\qquad\qquad\Lambda_{5}=-\frac{\epsilon^{2}}{2c_{s}^{4}}\left(4-\epsilon\right),\qquad\qquad\Lambda_{6}=\frac{\epsilon^{3}}{4c_{s}^{4}},\\
\Lambda_{4} & = & \Lambda_{7}=\Lambda_{8}=0.
\end{eqnarray*}
where $\eta_{\epsilon} = \frac{d\ln \epsilon}{d\ln a}$.
In \cite{Chen:2006nt}, the contribution to the bispectrum of curvature perturbation up to the second-order in slow-varying parameters were calculated, which correspond to our $\Lambda_1$, $\Lambda_2$, $\Lambda_3$ and $\Lambda_5$, whereas $\Lambda_6$ was neglected in \cite{Chen:2006nt} as it is third order in $\epsilon$.

In the last term in (\ref{S3}), we introduced
    \begin{equation}{\label{eom_linear}}
        \left.\frac{\delta{\cal L}_{2}}{\delta\zeta}\right|_{1} \equiv -2\left[\left(a^{2}\frac{\epsilon_s}{c_s^2}\zeta'\right)'-\epsilon_s a^2 \partial^{2}\zeta\right],
    \end{equation}
which is proportional to the linear equation of motion for $\zeta$ obtained by varying (\ref{S2}) with respect to $\zeta$, and
    \begin{equation}
        F(\zeta) = f_{1}\zeta\zeta'+f_{2}\left[\left(\partial_{i}\zeta\right)^{2}-\partial^{-2}\partial_{i}\partial_{j}\left(\partial^{i}\zeta\partial^{j}\zeta\right)\right]+f_{3}\left[\partial_{i}\zeta\partial^{i}\psi-\partial^{-2}\partial_{i}\partial_{j}\left(\partial^{i}\zeta\partial^{j}\psi\right)\right],
    \end{equation}
with
    \begin{eqnarray}
        f_{1} & = & -\frac{\epsilon_{T}}{a^{2}\mathcal{H}c_{s}^{2}c_{T}^{2}f_{\alpha}},\label{f1}\\
        f_{2} & = & \frac{1}{4a^{2}\mathcal{H}^{2}f_{\alpha}^{2}}\left(2\lambda_{1}f_{\alpha}+\lambda_{2}\frac{\epsilon_{T}}{c_{T}^{2}}\right),\label{f2}\\
        f_{3} & = & -\frac{1}{2a^{2}\mathcal{H}f_{\alpha}}\frac{\epsilon_{s}c_{T}^{4}}{\epsilon_{T}^{2}c_{s}^{2}}\left(\lambda_{1}f_{\alpha}+\frac{\epsilon_{T}}{c_{T}^{2}}\lambda_{2}\right).\label{f3}
        \end{eqnarray}

%There are three new types of three-point interactions in our generalized Galileon model, which disappear in $k$-essence model: terms proportional to $\Lambda_4$, $\Lambda_7$ and $\Lambda_8$ in (\ref{S3}). This provides us the possibility of being able to distinguish our model from $k$-essence model by detecting the different contributions to the bispectrum. However, as was shown recently in \cite{defe_ng}, these three terms can also arise if $f(\phi)R+\xi(\phi)\mathcal{G}$ terms are included, where $\mathcal{G}$ is the Gauss-Bonnet term. This is not surprising as the action we consider (\ref{action}) is the most general action with second order equations of motion, and hence includes the scalar-tensor model with $f(\phi)R+\xi(\phi)\mathcal{G}$ terms.

\subsection{Bispectrum of the curvature perturbation}

Having derived the full third order perturbation action (\ref{S3}), it is a standard but subtle\footnote{We have only calculated the bulk contributions to the third order action (\ref{S3}) and {\it not} kept track of the boundary terms. This should give the correct result for the bispectrum: see  \cite{Burrage:2011hd} and \cite{Arroja:2011yj} for a discussion of this point in the case of $k$-inflation.} exercise  \cite{Burrage:2011hd} to evaluate the corresponding bispectrum, i.e.~the three point function of curvature perturbation $\zeta$. The tree-level contributions to the three-point functions of $\zeta$ from cubic interactions described by (\ref{S3}) can be evaluated using
    \begin{equation}{\label{3pfun}}
        \left\langle \zeta_{\bm{k}_{1}} \left(\eta\right)\zeta_{\bm{k}_{2}} \left(\eta\right)\zeta_{\bm{k}_{3}}\left(\eta\right)\right\rangle =-2\Re\left[i\int_{\eta_{i}}^{\eta}d\eta'\left\langle 0\left|\zeta_{\bm{k}_{1}}\left(\eta\right)\zeta_{\bm{k}_{2}}\left(\eta\right)\zeta_{\bm{k}_{3}}\left(\eta\right)H_{\text{i}}\left(\eta'\right)\right|0\right\rangle \right],
    \end{equation}
where in the right-hand-side of (\ref{3pfun}) all quantities are in the interaction-picture, $\eta_i$ is some initial time when perturbation modes are deep inside the Hubble radius, which can be conveniently chosen to be $\eta_i = -\infty$. At cubic level, the Hamiltonian density is simply given by $\mathcal{H}_{(3)}= -\mathcal{L}_{(3)}$.

In the following, we assume that the eight dimensionless coefficients $\Lambda_i$ ($i=1,\cdots,8$) are approximately constant when evaluating the three point functions using (\ref{3pfun}). We simply collect the final results, and on denoting
    \begin{equation}
        \left\langle \zeta_{\bm{k}_{1}}\zeta_{\bm{k}_{2}}\zeta_{\bm{k}_{3}}\right\rangle =\left(2\pi\right)^{3}\delta^{3}\left(\bm{k}_{1}+\bm{k}_{2}+\bm{k}_{3}\right)B_{\zeta}\left(k_{1},k_{2},k_{3}\right)
    \end{equation}
the bispectrum $B_\zeta(k_1,k_2,k_3)$ is given by
    \begin{eqnarray}
B_{\zeta} & = & \frac{\left(2\pi^{2}\right)^{2}c_s^2{\cal P}_{\zeta}^{2}}{k_{1}^{3}k_{2}^{3}k_{3}^{3}\epsilon_s}\left[6\frac{(k_{1}k_{2}k_{3})^{2}}{K^{3}}\left(\Lambda_{1}+\frac{2}{c_{s}^{2}}\Lambda_{4}\right)+\frac{\Lambda_{2}}{K}\left(2\sum_{i>j}k_{i}^{2}k_{j}^{2}-\frac{1}{K}\sum_{i\neq j}k_{i}^{2}k_{j}^{3}\right)\right.\nonumber \\
 &  & \qquad+\frac{\Lambda_{3}}{2c_{s}^{2}}\left(\sum_{i}k_{i}^{3}+\frac{4}{K}\sum_{i>j}k_{i}^{2}k_{j}^{2}-\frac{2}{K^{2}}\sum_{i\neq j}k_{i}^{2}k_{j}^{3}\right)+\frac{\Lambda_{5}}{2}\left(\sum_{i}k_{i}^{3}-\frac{1}{2}\sum_{i\neq j}k_{i}k_{j}^{2}-\frac{2}{K^{2}}\sum_{i\neq j}k_{i}^{2}k_{j}^{3}\right)\nonumber \\
 &  & \qquad+\frac{\Lambda_{6}}{2K^{2}}\left(2\sum_{i}k_{i}^{5}+\sum_{i\neq j}k_{i}k_{j}^{4}-3\sum_{i\neq j}k_{i}^{2}k_{j}^{3}-2k_{1}k_{2}k_{3}\sum_{i>j}k_{i}k_{j}\right)\nonumber \\
 &  & \qquad+\frac{3}{2c_{s}^{4}}\frac{\Lambda_{7}}{K}\left(\sum_{i}k_{i}^{4}-2\sum_{i>j}k_{i}^{2}k_{j}^{2}\right)\left(1+\frac{1}{K^{2}}\sum_{i>j}k_{i}k_{j}+\frac{3k_{1}k_{2}k_{3}}{K^{3}}\right)\nonumber \\
 &  & \qquad\left.+\frac{1}{4c_{s}^{2}}\frac{\Lambda_{8}}{K^{2}}\left(7K\sum_{i}k_{i}^{4}+3k_{1}k_{2}k_{3}\sum_{i}k_{i}^{2}-2\sum_{i}k_{i}^{5}-5k_{1}k_{2}k_{3}K^{2}-12\sum_{i\neq j}k_{i}^{2}k_{j}^{3}\right)\right]\label{B_zeta}
\end{eqnarray}
where $K=k_1 + k_2 + k_3$.
%This equation is one of the main results of this paper.

From (\ref{B_zeta}), various contributions to the bispectrum $B_{\zeta}$ from different three-point interaction terms in (\ref{S3}) can be determined. Notice that $\zeta'^3$ and $\zeta'^2\partial^2\zeta$ contribute the same momentum shape. This can be understood as follows: since we are using the linear solution of $\zeta$ to evaluate the three-point function, we can  replace $\partial^2\zeta$ in $\zeta'^2\partial^2\zeta$ using the linear equation (\ref{eom_linear}), i.e. $c_{s}^{2}\partial^{2}\zeta=2\mathcal{H}\zeta'+\zeta''$ (note we are approximating $H$, $\epsilon_s$, $c_s$, $\Lambda_i$'s etc as constant). Thus we can write
    \[
        \frac{a^{2}}{\mathcal{H}^{2}}\Lambda_{4}\zeta'^{2}\partial^{2}\zeta=\frac{a^{2}}{\mathcal{H}^{2}}\Lambda_{4}\zeta'^{2}\frac{1}{c_{s}^{2}}\left(2\mathcal{H}\zeta'+\zeta''\right)\simeq\frac{a^{2}}{\mathcal{H}}\frac{2\Lambda_{4}}{c_{s}^{2}}\zeta'^{3},
    \]
which gives the correct coefficient $\frac{2\Lambda_{4}}{c_{s}^{2}}$ in the first term in (\ref{B_zeta}).  In fact, for a similar reason, the shapes proportional to $\Lambda_{7,8}$ appearing in (\ref{B_zeta}) are not new and must be a linear combination of the others\footnote{See \cite{SRP}. We thank S.Renaux-Petel pointing out this important point to us.}. Indeed, on
using the linear equation of motion one can also integrate by parts the last two terms in (\ref{S3}) which become
\begin{eqnarray*}
a^{2}\frac{\Lambda_{7}}{\mathcal{H}^{2}}\left[\partial^{2}\zeta\left(\partial_{i}\zeta\right)^{2}-\zeta\partial_{i}\partial_{j}\left(\partial_{i}\zeta\partial_{j}\zeta\right)\right] & \simeq & a^{2}\left[6\frac{\Lambda_{7}}{\mathcal{H}}\frac{1}{c_{s}^{4}}\zeta'^{3}-9\frac{\Lambda_{7}}{c_{s}^{4}}\zeta\zeta'^{2}+3\frac{\Lambda_{7}}{c_{s}^{2}}\zeta\left(\partial_{i}\zeta\right)^{2}\right],\\
a^{2}\frac{\Lambda_{8}}{\mathcal{H}}\left[\partial^{2}\zeta\partial_{i}\zeta\partial_{i}\psi-\zeta\partial_{i}\partial_{j}\left(\partial_{i}\zeta\partial_{j}\psi\right)\right] & \simeq & a^{2}\left[\frac{3}{2}\Lambda_{8}\frac{1}{c_{s}^{2}}\zeta\zeta'^{2}-\frac{1}{2}\Lambda_{8}\zeta\left(\partial_{i}\zeta\right)^{2}+\frac{1}{c_{s}^{2}}3\Lambda_{8}\zeta'\partial_{i}\zeta\partial_{i}\psi\right].
\end{eqnarray*}
Effectively this means that we can set $\Lambda_4$, $\Lambda_{7}$ and $\Lambda_8$ to zero in (\ref{B_zeta}) and replace $\Lambda_{1,\ldots,6}$ by
\begin{eqnarray*}
\tilde{\Lambda}_{1} & = & \Lambda_{1}+\frac{2\Lambda_{4}}{c_{s}^{2}}+\frac{6\Lambda_{7}}{c_{s}^{4}},\\
\tilde{\Lambda}_{2} & = & \Lambda_{2}-\frac{9\Lambda_{7}}{c_{s}^{4}}+\frac{3\Lambda_{8}}{2c_{s}^{2}},\\
\tilde{\Lambda}_{3} & = & \Lambda_{3}+\frac{3\Lambda_{7}}{c_{s}^{2}}-\frac{\Lambda_{8}}{2},\\
\tilde{\Lambda}_{5} & = & \Lambda_{5}+\frac{3\Lambda_{8}}{c_{s}^{2}}\, , \qquad \tilde{\Lambda}_{6} = \Lambda_6 \, .
\end{eqnarray*}

Finally, one may be interested in evaluating the dimensionless non-linear parameters, defined by \cite{wmap}
    \begin{equation}
        f_{\text{NL}}(k_1,k_2,k_3) \equiv \frac{5}{6}\frac{B_{\zeta}}{\left(2\pi^{2}\right)^{2}\mathcal{P}_{\zeta}^{2}}\frac{\prod_{i=1}^{3}k_{i}^{3}}{\sum_{i=1}^{3}k_{i}^{3}}.
    \end{equation}
To get some idea about the amplitude of the bispectrum, here we choose the equilateral momentum configuration $k_1=k_2=k_3$ and obtain
    \begin{eqnarray}
f_{\text{NL}}^{\text{(equil)}} & = & \frac{5}{81}\frac{c_{s}^{2}}{\epsilon_{s}}\left(\tilde{\Lambda}_{1}+6\tilde{\Lambda}_{2}+\frac{51}{4}\frac{\tilde{\Lambda}_{3}}{c_{s}^{2}}-3\tilde{\Lambda}_{5}-3\tilde{\Lambda}_{6}\right)\nonumber \\
 & = & \frac{5}{81}\frac{c_{s}^{2}}{\epsilon_{s}}\left(\Lambda_{1}+6\Lambda_{2}+\frac{51}{4}\frac{\Lambda_{3}}{c_{s}^{2}}+\frac{2}{c_{s}^{2}}\Lambda_{4}-3\Lambda_{5}-3\Lambda_{6}-\frac{39}{4}\frac{\Lambda_{7}}{c_{s}^{4}}-\frac{51}{8}\frac{\Lambda_{8}}{c_{s}^{2}}\right),\label{fnl}
\end{eqnarray}
where $\Lambda_i$'s are given in (\ref{Lambda1})-(\ref{Lambda8}).

%%%%%%%%%%%%%%%%%%%%%%%%%%%%%%%%%%%%%%%%%%%%%%%%%
\section{Conclusion}

In this paper we have studied cosmological perturbation theory in the ``generalized Galileon" model proposed in \cite{Deffayet:2011gz, horndeski}. In 4 space-time dimensions this is the most general scalar field theory whose Lagrangian contains derivatives up to second order while keeping the equations of motion which are second order and lower. Our model, which includes $k$-inflation as a special case, represents a large class of scalar field models which has not been investigated so far.

The present work is the first step in exploring the cosmological implications of ``generalized Galileon" models.
By determining the most generic second and third order actions for the curvature perturbation, we calculated the power spectra of scalar and tensor perturbations as well as the bispectrum of the curvature perturbation.
We found modifications of both the amplitude and propagation speed of tensor perturbations due to the presence of $\mathcal{L}^{(2)}$ and $\mathcal{L}^{(3)}$ Galileon terms.
Correspondingly, the tensor-to-scalar ratio is modified relative to $k$-essence (see also \cite{kyy2}). We have also showed that, although there are higher-order derivatives in the Galileon model relative to $k$-inflation, there are no new contributions to the bispectrum of the curvature perturbation.
In order to get a feel for the strength of the bispectrum, we evaluated the non-linear parameter $f_{\text{NL}}^{\text{equil}}$ for the equilateral configuration $k_1=k_2=k_3$.

In the future, the strength and shape of primordial non-gaussianities will be constrained both by CMB data as well as, independently, by galaxy clustering data.  Indeed, it is expected  \cite{LSS} that upcoming observations of high redshift clusters may possibly enable one to put limits on $f_{\text{NL}}$ at the level of a few tens. It remains a very interesting and open question to see whether, as a result of this different data, strong constraints may be put on the different unknown functions appearing in the generalized Galileon model. Our result (\ref{B_zeta}) will be a starting point for such an analysis.

%%%%%%%%%%%%%%%%%%%%%%%%%%%%%%%%%%%%%%%%%%%%%%%%%

\acknowledgments

It is a pleasure to thank
Antonio De Felice,
Guillaume Faye
for valuable discussions and kind help. We would particularly like to thank S\'ebastien Renaux-Petel for pointing out an important error in the first
version of this manuscript.
We would like to thank David Langlois for many useful discussions and for his interest at the early stages of this project.
XG was supported by ANR (Agence Nationale de la Recherche) grant ``STR-COSMO" ANR-09-BLAN-0157-01.

%%%%%%%%%%%%%%%%%%%%%%%%%%%%%%%%%%%%%%%%%%%%%%%%

\appendix

%%%%%%%%%%%%%%%%%%%%%%%%%%%%%%%%%%%%%%%
\section{Equations of motion}{\label{sec:eom3}}

The explicitly second order equations of motion for $\phi$ are obtained from (\ref{eofmJ}), and are given by
    \begin{equation}
        \sum_{n=0}^{3}\mathcal{E}^{(n)} = 0,
    \end{equation}
with
    \begin{eqnarray}
        \mathcal{E}^{(0)} & = & \square\phi K_{,X}^{}+K_{,XX}^{}\nabla_{\mu}X\nabla^{\mu}\phi-2XK_{,X\phi}^{}+K_{,\phi}^{},\label{eom0}\\
        \mathcal{E}^{(1)} & = & 2\square\phi G_{,\phi}^{(1)}-2XG_{,\phi\phi}^{(1)}+G_{,X}^{(1)}\left[\left(\square\phi\right)^{2}-\left(\nabla_{\mu}\nabla_{\nu}\phi\right){}^{2}-R_{\mu\nu}\nabla^{\mu}\phi\nabla^{\nu}\phi\right]\nonumber \\
         &  & -2G_{,X\phi}^{(1)}\left(X\square\phi-\nabla_{\mu}X\nabla^{\mu}\phi\right)+G_{,XX}^{(1)}\left(\nabla_{\mu}X\nabla^{\mu}X+\square\phi\nabla_{\mu}X\nabla^{\mu}\phi\right),\label{eom1}
        \end{eqnarray}
and
    \begin{eqnarray}
        \mathcal{E}^{(2)} & = & G_{,XX}^{(2)}\Big[\left(\square\phi\right)^{3}-3\square\phi\left(\nabla_{\mu}\nabla_{\nu}\phi\right)^{2}+2\left(\nabla_{\mu}\nabla_{\nu}\phi\right)^{3}-2\square\phi R_{\mu\nu}\nabla^{\mu}\phi\nabla^{\nu}\phi\nonumber \\
         &  & \qquad-4R_{\mu\nu}\nabla^{\mu}\phi\nabla^{\nu}X+R\nabla_{\mu}X\nabla^{\mu}\phi+2R_{\rho\mu\sigma\nu}\nabla^{\rho}\phi\nabla^{\sigma}\phi\nabla^{\mu}\nabla^{\nu}\phi\Big]\nonumber \\
         &  & +G_{,XXX}^{(2)}\left[2\square\phi\nabla_{\mu}X\nabla^{\mu}X+\left(\left(\square\phi\right)^{2}-\left(\nabla_{\mu}\nabla_{\nu}\phi\right)^{2}\right)\nabla_{\rho}X\nabla^{\rho}\phi-2\nabla^{\mu}X\nabla^{\nu}X\nabla_{\mu}\nabla_{\nu}\phi\right].\nonumber \\
         &  & -2G_{,X}^{(2)}G_{\mu\nu}\nabla^{\mu}\nabla^{\nu}\phi+2G_{,XX\phi}^{(2)}\left[X\left(\left(\nabla_{\mu}\nabla_{\nu}\phi\right){}^{2}-\left(\square\phi\right)^{2}\right)+2\nabla_{\mu}X\nabla^{\mu}X+2\square\phi\nabla_{\mu}X\nabla^{\mu}\phi\right]\nonumber \\
         &  & +RG_{,\phi}^{(2)}+G_{,X\phi}^{(2)}\left[3\left(\left(\square\phi\right)^{2}
         -\left(\nabla_{\mu}\nabla_{\nu}\phi\right){}^{2}\right)-2RX-4R_{\mu\nu}\nabla^{\mu}\phi\nabla^{\nu}\phi\right]
         +2G_{,X\phi\phi}^{(2)}\left(\nabla_{\mu}X\nabla^{\mu}\phi-2X\square\phi\right).\label{eom2}
        \end{eqnarray}
Though it is not particularly eluminating, the expression for $\mathcal{E}^{(3)}$ is given by
    \begin{equation}
         \mathcal{E}^{(3)} = F_{1}G_{,X}^{(3)}+F_{2}G_{,XX}^{(3)}+F_{3}G_{,XXX}^{(3)}+F_{4}G_{,XX\phi}^{(3)}+F_{5}G_{,X\phi}^{(3)}+F_{6}G_{,X\phi\phi}^{(3)}+F_{7}G_{,\phi}^{(3)}+F_{8}G_{,\phi\phi}^{(3)},
    \end{equation}
where $F_1,\cdots,F_8$ are scalar functions given by
    \begin{eqnarray}
F_{1} & = & 3R\left[\left(\square\phi\right)^{2}-\left(\nabla_{\mu}\nabla_{\nu}\phi\right)^{2}-R_{\mu\nu}\nabla^{\mu}\phi\nabla^{\nu}\phi\right]+3\left(2R_{\mu\rho}R_{\nu}^{\rho}+2R^{\rho\sigma}R_{\mu\rho\nu\sigma}-R_{\mu}^{\phantom{\mu}\rho\sigma\lambda}R_{\nu\rho\sigma\lambda}\right)\nabla^{\mu}\phi\nabla^{\nu}\phi\nonumber \\
 &  & +12R_{\mu\nu}\left(\nabla^{\rho}\nabla^{\mu}\phi\nabla_{\rho}\nabla^{\nu}\phi-\square\phi\nabla^{\mu}\nabla^{\nu}\phi\right)+6R_{\mu\rho\nu\sigma}\nabla^{\mu}\nabla^{\nu}\phi\nabla^{\rho}\nabla^{\sigma}\phi,\label{F1}\\
F_{2} & = & 3R\left(\nabla_{\mu}X\nabla^{\mu}X+\square\phi\nabla_{\mu}X\nabla^{\mu}\phi\right)-3\left[\left(\square\phi\right)^{2}-\left(\nabla_{\rho}\nabla_{\sigma}\phi\right)^{2}\right]R_{\mu\nu}\nabla^{\mu}\phi\nabla^{\nu}\phi\nonumber \\
 &  & +6R_{\mu\nu}\left(2\nabla^{\mu}\phi\nabla^{\nu}\nabla_{\rho}\phi\nabla^{\rho}X-\nabla^{\mu}X\nabla^{\nu}X-\nabla^{\mu}\nabla^{\nu}\phi\nabla_{\rho}X\nabla^{\rho}\phi-2\square\phi\nabla^{\mu}X\nabla^{\nu}\phi\right)\nonumber \\
 &  & +6R_{\rho\mu\sigma\nu}\left[\left(\square\phi\nabla^{\mu}\nabla^{\nu}\phi-\nabla^{\mu}\nabla^{\lambda}\phi\nabla^{\nu}\nabla_{\lambda}\phi\right)\nabla^{\rho}\phi\nabla^{\sigma}\phi+2\nabla^{\rho}\phi\nabla^{\sigma}X\nabla^{\mu}\nabla^{\nu}\phi\right],\label{F2}\\
F_{3} & = & 3\left[\left(\square\phi\right)^{2}-\left(\nabla_{\mu}\nabla_{\nu}\phi\right)^{2}\right]\nabla_{\rho}X\nabla^{\rho}X+6\nabla^{\mu}X\nabla^{\nu}X\left(\nabla_{\mu}\nabla^{\rho}\phi\nabla_{\rho}\nabla_{\nu}\phi-\square\phi\nabla_{\mu}\nabla_{\nu}\phi\right)\nonumber \\
 &  & +\left[\left(\square\phi\right)^{3}-3\square\phi\left(\nabla_{\mu}\nabla_{\nu}\phi\right)^{2}+2\left(\nabla_{\mu}\nabla_{\nu}\phi\right)^{3}\right]\nabla_{\rho}X\nabla^{\rho}\phi,\label{F3}\\
F_{4} & = & 12\left(\square\phi g_{\mu\nu}-\nabla_{\mu}\nabla_{\nu}\phi\right)\nabla^{\mu}X\nabla^{\nu}X+6\left[\left(\square\phi\right)^{2}-\left(\nabla_{\mu}\nabla_{\nu}\phi\right)^{2}\right]\nabla_{\rho}X\nabla^{\rho}\phi\nonumber \\
 &  & -2X\left[\left(\square\phi\right)^{3}-3\square\phi\left(\nabla_{\mu}\nabla_{\nu}\phi\right)^{2}+2\left(\nabla_{\mu}\nabla_{\nu}\phi\right)^{3}\right],\label{F4}\\
F_{5} & = & 4\left[\left(\square\phi\right)^{3}-3\square\phi\left(\nabla_{\mu}\nabla_{\nu}\phi\right)^{2}+2\left(\nabla_{\mu}\nabla_{\nu}\phi\right)^{3}\right]-6R\left(X\square\phi-\nabla^{\mu}X\nabla_{\mu}\phi\right)\nonumber \\
 &  & +12R_{\mu\nu}\left[X\nabla^{\mu}\nabla^{\nu}\phi-\square\phi\nabla^{\mu}\phi\nabla^{\nu}\phi-2\nabla^{\mu}\phi\nabla^{\nu}X\right]+12R_{\rho\beta\alpha\mu}\nabla^{\rho}\phi\nabla^{\alpha}\phi\nabla^{\mu}\nabla^{\beta}\phi,\label{F5}\\
F_{6} & = & -6X\left[\left(\square\phi\right)^{2}-\left(\nabla_{\mu}\nabla_{\nu}\phi\right)^{2}\right]+6\nabla_{\mu}X\nabla^{\mu}X+6\square\phi\nabla_{\mu}X\nabla^{\mu}\phi,\label{F6}\\
F_{7} & = & -12G_{\mu\nu}\nabla^{\mu}\nabla^{\nu}\phi,\label{F7}\\
F_{8} & = & -6G_{\mu\nu}\nabla^{\mu}\phi\nabla^{\nu}\phi.\label{F8}
\end{eqnarray}

%%%%%%%%%%%%%%%%%%%%%%%%%%%%%%%%%%%%%%%%
\section{Explicit expression for $T^{(3)}_{\mu\nu}$}{\label{sec:EMT3}}

The energy-momentum tensor corresponding to $\mathcal{L}^{(3)}$ takes the following form:
    \begin{eqnarray}
        T_{\mu\nu}^{(3)} & = & C_{1}g_{\mu\nu}+C_{2}\nabla_{\mu}\phi\nabla_{\nu}\phi+C_{3}\nabla_{\mu}X\nabla_{\nu}X+C_{4}\nabla_{(\mu}\phi\nabla_{\nu)}X+C_{5}\nabla_{\mu}\nabla_{\nu}\phi+C_{6}\nabla_{\beta}X\nabla^{\beta}\nabla_{(\nu}\phi\nabla_{\mu)}\phi\nonumber \\
         &  & +C_{7}\text{ }\nabla^{\text{\ensuremath{\beta}1}}X\nabla_{\text{\ensuremath{\beta}1}}\nabla_{\beta}\phi\nabla^{\beta}\nabla_{(\mu}\phi\nabla_{\nu)}\phi+C_{8}\nabla^{\alpha}\nabla_{\mu}\phi\nabla_{\nu}\nabla_{\alpha}\phi+C_{9}\nabla_{\text{\ensuremath{\beta}1}}X\nabla_{(\nu}X\nabla^{\text{\ensuremath{\beta}1}}\nabla_{\mu)}\phi\nonumber \\
         &  & +C_{10}\nabla_{\beta}\nabla_{\alpha}\phi\nabla^{\alpha}\nabla_{\mu}\phi\nabla^{\beta}\nabla_{\nu}\phi+\tau_{\mu\nu}^{(3)},\label{EMT3}
        \end{eqnarray}
where $C_1,\cdots , C_{10}$ are scalar coefficients which are given by
    \begin{eqnarray}
C_{1} & = & 6G_{,\phi}^{(3)}\left[\left(\nabla_{\rho}\nabla_{\sigma}\phi\right)^{2}-\left(\square\phi\right)^{2}+2G_{\rho\sigma}\nabla^{\rho}\phi\nabla^{\sigma}\phi-RX\right]-6G_{,\phi\phi}^{(3)}\left(\nabla_{\rho}X\nabla^{\rho}\phi-2X\square\phi\right)\nonumber \\
 &  & -6G_{,X\phi}^{(3)}\left(X\left(\left(\nabla_{\rho}\nabla_{\sigma}\phi\right){}^{2}-\left(\square\phi\right)^{2}\right)+2\nabla_{\rho}X\nabla^{\rho}X+2\square\phi\nabla_{\rho}X\nabla^{\rho}\phi\right)\nonumber \\
 &  & -3G_{,X}^{(3)}\Big[\frac{2}{3}\left(\left(\Box\phi\right)^{3}-3\square\phi\left(\nabla_{\rho}\nabla_{\sigma}\phi\right){}^{2}+2\left(\nabla_{\rho}\nabla_{\sigma}\phi\right)^{3}\right)-2\square\phi R_{\rho\sigma}\nabla^{\rho}\phi\nabla^{\sigma}\phi\nonumber \\
 &  & -4R_{\rho\sigma}\nabla^{\rho}X\nabla^{\sigma}\phi+R\nabla_{\rho}X\nabla^{\rho}\phi+2R_{\rho\lambda\sigma\tau}\nabla^{\rho}\phi\nabla^{\sigma}\phi\nabla^{\lambda}\nabla^{\tau}\phi\Big]\nonumber \\
 &  & -G_{,XX}^{(3)}\left[3\left(\left(\square\phi\right)^{2}-\left(\nabla_{\rho}\nabla_{\sigma}\phi\right){}^{2}\right)\nabla_{\lambda}X\nabla^{\lambda}\phi-6\nabla^{\rho}X\nabla^{\sigma}X\nabla_{\rho}\nabla_{\sigma}\phi+6\square\phi\nabla^{\rho}X\nabla_{\rho}X\right],\label{C1}
\end{eqnarray}
\begin{eqnarray}
C_{2} & = & 6G_{,X\phi}^{(3)}\left[\left(\square\phi\right)^{2}-\left(\nabla_{\rho}\nabla_{\sigma}\phi\right){}^{2}\right]+6G_{,\phi}^{(3)}R+6G_{,\phi\phi}^{(3)}\square\phi-6G_{,X}^{(3)}G_{\rho\sigma}\nabla^{\rho}\nabla^{\sigma}\phi\nonumber \\
 &  & +G_{,XX}^{(3)}\left[\left(\square\phi\right)^{3}-3\left(\nabla_{\rho}\nabla_{\sigma}\phi\right)^{2}\square\phi+2\left(\nabla_{\rho}\nabla_{\sigma}\phi\right)^{3}\right],\label{C2}\\
C_{3} & = & 12G_{,X\phi}^{(3)}+6G_{,XX}^{(3)}\square\phi,\label{C3}\\
C_{4} & = & 6G_{,X}^{(3)}R+6G_{,XX}^{(3)}\left[\left(\square\phi\right)^{2}-\left(\nabla_{\rho}\nabla_{\sigma}\phi\right)^{2}\right]+24G_{,X\phi}^{(3)}\square\phi+12G_{,\phi\phi}^{(3)},\label{C4}\\
C_{5} & = & 12G_{,\phi}^{(3)}\square\phi-12G_{,\phi\phi}^{(3)}X-6G_{,X}^{(3)}\left[\left(\nabla_{\rho}\nabla_{\sigma}\phi\right)^{2}-\left(\square\phi\right)^{2}+R_{\rho\sigma}\nabla^{\rho}\phi\nabla^{\sigma}\phi\right]\nonumber \\
 &  & -2G_{,X\phi}^{(3)}\left(6X\square\phi-6\nabla_{\rho}X\nabla^{\rho}\phi\right)+6G_{,XX}^{(3)}\left(\nabla_{\rho}X\nabla^{\rho}X+\square\phi\nabla_{\rho}X\nabla^{\rho}\phi\right),\label{C5}\\
C_{6} & = & -12G_{,XX}^{(3)}\square\phi-24G_{,X\phi}^{(3)},\label{C6}\\
C_{7} & = & 12G_{,XX}^{(3)},\label{C7}\\
C_{8} & = & -12G_{,X}^{(3)}\square\phi+12G_{,X\phi}^{(3)}X-12G_{,\phi}^{(3)}-6G_{,XX}^{(3)}\nabla_{a}X\nabla^{a}\phi,\label{C8}\\
C_{9} & = & -12G_{,XX}^{(3)},\label{C9}\\
C_{10} & = & 12G_{,X}^{(3)},\label{C10}
\end{eqnarray}
and $\tau_{\mu\nu}^{(3)}$ represents terms whose $\mu,\nu$-indices explicitly depend on or coupled to Riemannian tensors:
\begin{eqnarray}
\tau_{\mu\nu}^{(3)} & = & -12G_{,\phi}^{(3)}\left(XR_{\mu\nu}+R_{\mu\rho\nu\sigma}\nabla^{\rho}\phi\nabla^{\sigma}\phi+2\nabla_{(\mu}\phi R_{\nu)\rho}\nabla^{\rho}\phi\right)\nonumber \\
 &  & -6G_{,X}^{(3)}\Big\{2\left[\left(\square\phi\nabla^{\rho}\phi+\nabla^{\rho}X\right)R_{\rho(\mu}+\nabla^{\sigma}\nabla^{\rho}\phi\nabla^{\lambda}\phi R_{\lambda\rho\sigma(\mu}-R_{\rho\sigma}\nabla^{\rho}\phi\nabla^{\sigma}\nabla_{(\mu}\phi\right]\nabla_{\nu)}\phi\nonumber \\
 &  & \qquad\qquad-\nabla_{\rho}X\nabla^{\rho}\phi R_{\mu\nu}+2\nabla^{\rho}\phi R_{\rho(\mu}\nabla_{\nu)}X\nonumber \\
 &  & \qquad\qquad-R_{\rho(\mu\nu)\sigma}\nabla^{\rho}\phi\left(\square\phi\nabla^{\sigma}\phi+2\nabla^{\sigma}X\right)-2\nabla^{\rho}\phi\nabla^{\lambda}\phi R_{\lambda\sigma\rho(\mu}\nabla^{\sigma}\nabla_{\nu)}\phi\Big\}.\label{tau3}
\end{eqnarray}

%%%%%%%%%%%%%%%%%%%%%%%%%%%%%%%%
\section{General consideration of the structure of perturbative action}{\label{sec:cancel}}

Due to the complexity of the perturbative calculation, it is useful try to deduce some general properties of the perturbed action.  As we now show, on using the background equations of motion, two types of cancellation occur rather generally. As a result the conclusions of this appendix are useful not only to simplify expressions but also provide us consistency check in practical calculations.

%\subsection{Absence of $\zeta^n$, $\alpha\zeta^{n-1}$ terms in the $n$-th order perturbative action}

We first argue that in the quadratic Lagrangian there are no ``mass term" $\zeta^2$ and $\alpha\zeta$; and similarly in the cubic Lagrangian there are no $\zeta^3$ and $\alpha\zeta^2$. To this end, we focus on the corresponding fully non-perturbative Lagrangian of (\ref{action}), substitute (\ref{perturbedm}), and furthermore neglect all spatial gradient terms. This yields
    \begin{equation}{\label{L_ls_gg}}
        \mathcal{L} \approx e^{3\zeta}\left[ A(\zeta',\alpha) + \alpha' B(\zeta',\alpha) +\zeta'' C(\zeta',\alpha) \right]\,,
    \end{equation}
with
    \begin{eqnarray}
        A&=&\frac{1}{a^{2}}e^{-5\alpha}\Big[a^{6}e^{6\alpha}K^{}+6G_{,X}^{(3)}(\mathcal{H}+\zeta')^{2}
        \left(\phi'\right)^{2}\left((2\mathcal{H}-\zeta')\phi'-3\phi''\right)+\nonumber\\
        &&a^{4}e^{4\alpha}\left(3(\mathcal{H}+\zeta')(\mathcal{H}+2\zeta')+3\mathcal{H}'
        +6G^{(2)}\left((\mathcal{H}+\zeta')(\mathcal{H}+2\zeta')+\mathcal{H}'\right)-G^{(1)}\left((2\mathcal{H}+3\zeta')\phi'
        +\phi''\right)\right) \nonumber\\
        &&-6a^{2}e^{2\alpha}(\mathcal{H}+\zeta')\left(-G_{,X}^{(2)}\phi'\left(\zeta'\phi'+\phi''\right)
        +3G^{(3)}\left(\left(3\zeta'(\mathcal{H}+\zeta')
        +2\mathcal{H}'\right)\phi'+(\mathcal{H}+\zeta')\phi''\right)\right)\Big],\label{A_ls}\\
        B&=&\frac{1}{a^{2}}e^{-5\alpha}\Big[18G_{,X}^{(3)}(\mathcal{H}+\zeta')^{2}\left(\phi'\right)^{3}-a^{4}e^{4\alpha}
        \left(3\left(1+2G^{(2)}\right)(\mathcal{H}+\zeta')-G_{,X}^{(1)}\phi'\right)\nonumber\\
        &&+6a^{2}e^{2\alpha}(\mathcal{H}+\zeta')\phi'\left(9G^{(3)}(\mathcal{H}+\zeta')
        -G_{,X}^{(2)}\phi'\right)\Big],\label{B_ls}\\
        C&=&e^{-3\alpha}\left(3a^{2}e^{2\alpha}\left(1+2G^{(2)}\right)-36G^{(3)}(\mathcal{H}+\zeta')\phi'\right).\label{C_ls}
\end{eqnarray}
We emphasize that in (\ref{A_ls})-(\ref{C_ls}) we neglect all spatial derivatives in the functions $K$ and $G^{(1,2,3)}$, so for example $K^{} = K^{}(\frac{1}{2}\frac{e^{-2\alpha}}{a^{2}}\phi'^{2},\phi)$. Notice that $\zeta$ enters the Lagrangian (\ref{L_ls_gg}) only through an overall prefactor $e^{3\zeta}$.

Although (\ref{L_ls_gg}) can further simplified through integration-by-parts, it is already adequate for the following discussion. The perturbative expansion of (\ref{L_ls_gg}) yields
    \begin{equation}
        \mathcal{L}=\mathcal{L}_{\left(0\right)}+\mathcal{L}_{\left(1\right)}+\mathcal{L}_{\left(2\right)}+\mathcal{L}_{\left(3\right)}+\dots,
    \end{equation}
where the linear action is
    \begin{equation}
        \mathcal{L}_{\left(1\right)} = \left(3A-A_{,\zeta'}'+C''\right)\zeta+\left(A_{,\alpha}-B'\right)\alpha,
    \end{equation}
which is vanishing and yields the corresponding background equations of motion
    \begin{eqnarray}
        \bar{\mathcal{E}}_{\zeta} & \equiv & 3A-A_{,\zeta'}'+C''=0,\label{bgeom_zeta}\\
        \bar{\mathcal{E}}_{\alpha} & \equiv & A_{,\alpha}-B'=0,\label{bgeom_alpha}
        \end{eqnarray}
where quantities are understood as their background values.
After integration-by-parts, in the quadratic Lagrangian terms which which are proportional to $\zeta^2$ and $\zeta\alpha$ are
    \begin{equation}{\label{L_ls_2}}
        \mathcal{L}_{\left(2\right)} \supset \frac{3}{2}\left(3A-A_{,\zeta'}'+C''\right)\zeta^{2}+3\left(A_{,\alpha}-B'\right)\zeta\alpha,
    \end{equation}
and in the cubic Lagrangian:
    \begin{equation}{\label{L_ls_3}}
        \mathcal{L}_{\left(3\right)} \supset  \frac{3}{2}\left(3A-A_{,\zeta'}'+C''\right)\zeta^{3}+\frac{9}{2}\left(A_{,\alpha}-B'\right)\zeta^{2}\alpha.
        \end{equation}
At this point, it is explicit that $\zeta^2$ and $\alpha\zeta$ in the quadratic Lagrangian (\ref{L_ls_2}), $\zeta^3$ and $\alpha\zeta^2$ in the cubic Lagrangian (\ref{L_ls_3}) are proportional to the background equations (\ref{bgeom_zeta})-(\ref{bgeom_alpha}) respectively, which are thus vanishing.

The above analysis can be directly generalized to the higher orders. Generally, in $n$-th order perturbative action, $\zeta^n$ the proportional to the background equation of $\zeta$ (\ref{bgeom_zeta}),  $\alpha\zeta^{n-1}$ is proportional to the background equation of $\alpha$ (\ref{bgeom_alpha}), which is the background energy constraint.
Thus both $\zeta^n$ and $\alpha\zeta^{n-1}$ must be vanishing in the $n$-th order perturbative action{\footnote{It is the absence of self-interaction term $\zeta^n$ that ensures the conservation of curvature perturbation $\zeta$ on large scales and on fully-nonlinear order in the model (\ref{action}). See \cite{Gao:2011mz} for details.}}.

%\subsection{Absence of self-interaction terms of $\beta$}

Another type of cancelations is the self-interaction terms of $\beta$. One can explicitly check that the coefficient before $\beta^3$ self-interaction terms is proportional to the background energy constraint and thus vanishes.
%This can easily understood by observing that we are working with ADM variables $N=ae^{\alpha}$, $N_i =a^2\partial_{i}\beta$ and $h_{ij} =a^{2}e^{2\zeta}\delta_{ij}$, thus
%{\gao{talk about something}}

We wish to emphasize that the above cancellation is a virtue of choosing ADM-compatible perturbation variables{\footnote{We would like to thank Antonio De Felice for pointing out this.}}, i.e. $N=ae^{\alpha}$ and $N=a^2\partial_i\beta$. If one alternatively chooses other variables, in general there is no such cancellation.

%
%\section{Explicit expressions for $\epsilon_s$ and $c_s^2$}{\label{sec:gamma_cs}}
%{\gao{remove this appendix???}}
%The normalization factor $\epsilon_s$ and speed of sound $c_s^2$ in (\ref{S2}) are given by

%%%%%%%%%%%%%%%%%%%%%%%%%%%%%%%%%%%%%%%%%%%%%%%%%%%%%%%%%%%%%%%%%%
%%********************  Bibliography.Begin  ********************%%

%%********************  Bibliography.End  ********************%%

\end{document}